\documentclass[12pt]{JHEP}

\usepackage{latexsym,amsmath,amsfonts,amssymb,epsfig}

\def\be{\begin{equation}}
\def\ee{\end{equation}}
\def\bd{\begin{displaymath}}
\def\ed{\end{displaymath}}

\def\lp{\left(}
\def\rp{\right)}
\def\ba{\begin{eqnarray}}
\def\ea{\end{eqnarray}}

\def\n1{${\cal N}=1$}

\def\tr{\ensuremath{\mathrm {Tr\:}}}
\def\c={\ensuremath{\overset{c.r.}{=}}}

\def\pk1{\ensuremath{\frac{{\cal W}_{\alpha}^{(k+1)}{\cal W}^{(k+1)\alpha}}{z-\Phi_{(k+1)}}}}
\def\p1{\ensuremath{\frac{{\cal W}_{\alpha}^{(1)}{\cal W}^{(1)\alpha}}{z-\Phi_{(1)}}}}

\def\pW2{\ensuremath{\frac{{\cal W}_{\alpha}^{(2)}{\cal W}^{(2)\alpha}}{z-\Phi_{(2)}}}}
\def\X{\ensuremath{X_{k,k+1}}}
\def\Xb{\ensuremath{\bar{X}_{k+1,k}}}
\def\i{\ensuremath{\bar{\imath}}}
\def\j{\ensuremath{\bar{\jmath}}}
\def\pk{\ensuremath{\hat{\Phi}_k}}
\def\xk{\ensuremath{\hat{X}_k}}
\def\xbk{\ensuremath{\hat{\bar{X}}_k}}
\def\s{\ensuremath{\hat{S}_k}}
\def\g{\ensuremath{\frac{1}{g_{\scriptscriptstyle{S}}}}}
\def\gq{\ensuremath{\frac{1}{g_{\scriptscriptstyle{S}}^{\scriptscriptstyle{2}}}}}

\title{Phases and geometry of the $\mathcal{N}=1$ $A_2$ quiver gauge theory and matrix models}

\author{Roberto Casero and Enrico Trincherini \\ Dipartimento di Fisica\\Universit\`{a} di Milano - Bicocca\\ Piazza della Scienza, 3\\ 20126 Milano, Italy\\  \email{roberto.casero@mib.infn.it}, \email{enrico.trincherini@mib.infn.it}}

\abstract{We study the phases and geometry of the $\mathcal{N}=1$ $A_2$ quiver gauge theory using matrix models and a generalized Konishi anomaly. We consider the theory both in the Coulomb and Higgs phases. Solving the anomaly equations, we find that a meromorphic one-form $\sigma(z)dz$ is naturally defined on the curve $\Sigma$ associated to the theory. Using the Dijkgraaf-Vafa conjecture, we evaluate the effective low-energy superpotential and demonstrate that its equations of motion can be translated into a geometric property of $\Sigma$: $\sigma(z)dz$ has  integer periods around all compact cycles. This ensures that there exists on $\Sigma$ a meromorphic function whose logarithm $\sigma(z)dz$ is the differential. We argue that the surface determined by this function is the $\mathcal{N}=2$ Seiberg-Witten curve of the theory.}

\preprint{Bicocca-FT-03-19\\hep-th/0307054}

\begin{document}

\section{Introduction}

A new tool to study the low-energy dynamics of $\mathcal{N}=1$ supersymmetric gauge theories was introduced in \cite{DV}, inspired by the study of the geometric engineering of such theories into configurations of D-branes wrapped around non-trivial cycles of Calabi-Yau spaces, and of topological string theories \cite{{Bershadsky:1993cx},{Cachazo:2001jy},{Cach},{Dijkgraaf:2002fc},{Dijkgraaf:2002vw}}.

In \cite{DV} it was postulated that an $\mathcal{N}=1$  supersymmetric gauge theory is associated  to a matrix model built by replacing all superfields with bosonic matrices. It was argued that the effective low-energy superpotential of such field theories could be evaluated as a functional of the free energy of the associated matrix model. The superpotential that is obtained this way is a function of the low-energy glueball superfields $S_i$, which have as their lowest components the gluino bilinears of the confining factors $SU(N_i)$ of the low-energy gauge group. This relation is actually evaluated off-shell. The correct vacuum is found by minimizing the superpotential. The matrix model also naturally defines an algebraic curve which is encoded into its loop equations. 

For the $U(N)$ theory with matter in the adjoint representation, the conjecture was later studied and tested through purely field theoretic methods, but with completely different approaches, in  \cite{Ferrari1}, \cite{Grisaru} and \cite{CSW1}. In \cite{Ferrari1} a proof of the conjecture was given in the particular case of confining vacua. In \cite{Grisaru} and \cite{CSW1}, on the other side, a derivation of the perturbative part of the effective superpotential was given, which matches the matrix model calculation of \cite{DV}. In \cite{CSW1}, in particular,  use  was made of a generalized Konishi anomaly \cite{Konishi}, which allows to write non-perturbative relations for the resolvent of the gauge theory which are really closely related to the matrix model loop equations. In this framework, the expectation values for the glueball superfields may be expressed as the periods of the resolvent around a set of compact cycles of the algebraic curve associated to the gauge theory.

Many papers appeared recently, generalizing the work of \cite{DV} and \cite{CSW1} to more general gauge theories, and showing that the matrix model conjecture is valid in a wide range of cases \cite{partial}.

It looked like a puzzle how to translate into a geometric language the minimization problem, since the curve that was obtained through loop equations/Ward identities seemed to know nothing about the vacuum selection. This problem was solved in \cite{Ferrari2} and \cite{CSW3} where it was demonstrated that the equations of motion of the effective superpotential for a theory with matter in the adjoint only, and in the adjoint and fundamental representations, respectively,  are equivalent to imposing that a particular meromorphic one-form, satisfying another anomaly equation, has integer periods around every compact cycle of the algebraic curve.

In this paper we extend the work of \cite{CSW3} to the case of an $A_2$ quiver gauge theory. Quiver gauge theories  were introduced in \cite{DM} to describe the low-energy dynamics of branes placed at singular points of orbifold space times. The theory we consider is the simplest example of such theories. It has gauge group $U(N_1)\times U(N_2)$, and its matter content is made up of two chiral superfields $\Phi_1$ and $\Phi_2$, each in the adjoint of one of the two factors of the gauge group, and an $\mathcal{N}=2$ hypermultiplet $(X,\bar{X})$ in the bifundamental representation. The supersymmetry is broken from $\mathcal{N}=2$ to $\mathcal{N}=1$ by adding a superpotential for each of the adjoint superfields. The tree superpotential is given by
\be
W_{tree}=\tr W_1(\Phi_1)+\tr W_2(\Phi_2)+\tr\lp \bar{X}\Phi_1 X-X\Phi_2\bar{X}\rp
\ee
where
\be
\begin{split}
&W_1^\prime(z)=g_{1,n_1+1}\prod_{i=1}^{n_1}(z-a_i)\\
&W_2^\prime(z)=g_{2,n_2+1}\prod_{j=1}^{n_2}(z-b_j)\\
&W_1^\prime(z)+W_2^\prime(z)=g_{1,n_1+1}\prod_{k=1}^{n_1}(z-c_k)
\end{split}
\ee

The quantities to be studied to give a description of the low-energy theory are \cite{CSW1}
\be\label{operators}
\begin{split}
&T_a(z)\equiv \tr \frac{1}{z-\Phi_a}\\
&w_a^\alpha(z)\equiv\frac{1}{4\pi}\tr\frac{\mathcal{W}_a^\alpha}{z-\Phi_a}\\
&R_a(z)\equiv -\frac{1}{32\pi^2}\tr\frac{\mathcal{W}_{a\alpha}\mathcal{W}_a^\alpha}{z-\Phi_a}
\end{split}
\ee

In \cite{{naculich},{quiver}} it was shown that the anomaly equations for the resolvents $R_a(z)$ define a non hyperelliptic curve $\Sigma$
\be
y^3- p_2(z)y - p_3(z)=0
\ee
where $p_2(z)$ and $p_3(z)$ are polynomials depending on the superpotentials $W_1(z)$ and $W_2(z)$, whose expression we report in section \ref{geom}. The roots of this equation are given by $R_1(z)$, $R_2(z)$ and $-R_1(z)-R_2(z)$.

The chiral ring operators \eqref{operators} are useful to write in a compact form the low energy degrees of freedom
\be\label{fielddegrees}
\begin{split}
&S_{1,i}=\frac{1}{2\pi i}\oint_{\tilde{A}_i}dz ~R_1(z)=-\frac{1}{2\pi i}\oint_{A_i}dz ~y(z)\\
&S_{2,j}=\frac{1}{2\pi i}\oint_{\tilde{B}_j}dz ~R_2(z)=\frac{1}{2\pi i}\oint_{B_j}dz ~y(z)\\
&H_{1,k}=\frac{1}{2\pi i}\oint_{C_k}dz ~R_1(z)=\frac{1}{2\pi i}\oint_{C_k}dz ~y(z)\\
&H_{2,k}=\frac{1}{2\pi i}\oint_{\tilde{C}_k}dz ~R_2(z)=\frac{1}{2\pi i}\oint_{C_k}dz ~y(z)
\end{split}
\ee
where $A_i$, $B_j$ and $C_k$ are compact cycles which encircle the critical points $a_i$, $b_j$ and $c_k$ respectively.

In this paper we write the effective superpotential for the $\mathcal{N}=1$ $A_2$ theory and show that its minimization is translated into geometric language in the condition that a meromorphic one-form $\sigma(z)dz$, closely related to the chiral ring operators $T_1(z)$ and $T_2(z)$, has integer periods around all compact cycles of $\Sigma$.

This condition ensures that a meromorphic function $\psi(z)$ can be defined on $\Sigma$ such that $\sigma(z)dz=d\,\ln \psi(z)$. This function defines a new surface which we argue to be identical to the Seiberg-Witten curve for the $\mathcal{N}=2$ theory.

The paper is organized as follows. In section 2 we describe the classical phase structure of the theory. In section 3 we describe the curve $\Sigma$ which is associated with the theory through the anomaly equations satisfied by the resolvents, and solve the Ward identities for the operators $T_1(z)$ and $T_2(z)$. This solution suggests that on $\Sigma$, a meromorphic one-form $\sigma(z)dz$ is naturally defined. In section 4 we study the matrix model associated with non-affine quiver gauge theories and evaluate their effective superpotential. In section 5 we describe the minimization of the effective superpotential from a geometric perspective and find the meromorphic function $\psi(z)$. In section 6 we comment on the results and on the connection between the different phases of the theory. At the end of the paper there are three appendices in which we study the holomorphic one-forms on $\Sigma$, report some computations of the matrix model and report the bilinear Riemann relations which we used in the paper.

\section{Classical phases of quivers}\label{phases}

In this section we study the classical phases of a generic non-affine quiver theory, and will specialize to the $A_2$ case only at the end. The superpotential of the theory is given by \cite{DM}
\be\label{Wtree}
W_{tree}=\sum_{k=1}^n\tr W_k(\Phi_k)+\sum_{k=1}^{n-1}\tr\lp \bar{X}_{k+1,k}\Phi_k X_{k,k+1}-X_{k,k+1}\Phi_{k+1}\bar{X}_{k+1,k} \rp
\ee

where
\be\label{chpotential}
W_k(\Phi_k)=\sum_{p=1}^{n_k+1}\frac{g_{k,p}}{p}\Phi_k^p
\ee

We label the critical points of the potentials $W_k(z)$ as $a_{k,i}$ $(i=1,\ldots n_k)$, so that
\be
W_k^\prime(z)=g_{k,n_k+1}\prod_{i=1}^{n_k}
(z-a_{k,i})\ee

There are two kinds of vacua for theories with matter in the adjoint and in the bifundamental representations of the gauge group. Coulomb vacua are characterized by zero classical expectation values for all the bifundamental fields, while Higgs vacua have non-zero VEV for some of them.

\subsection{Equations of motion}

We first write the equations of motion that are obtained by differentiating (\ref{Wtree}) with respect to the bifundamentals \X\, and \Xb
\be\label{eqmotbif}
\begin{split}
&\Xb\Phi_k=\Phi_{k+1}\Xb\\
&\X\Phi_{k+1}=\Phi_k\X
\end{split}
\ee

From these it is easy to obtain the two following commutation relations
\be\label{combif}
\begin{split}
&[\X\Xb,\Phi_k]=0\\
&[\Xb\X,\Phi_{k+1}]=0
\end{split}
\ee
where the two combinations \X\Xb~ and \Xb\X~ are in the adjoint representations of $U(N_k)$ and $U(N_{k+1})$ respectively. This allows us to simultaneously diagonalize $\Phi_k$ and \X\Xb~ and $\Phi_{k+1}$ and \Xb\X.

We can also use (\ref{eqmotbif}) to write (\ref{Wtree}) in a way that makes it more convenient to derive the equations of motion for the adjoint fields:
\be
W_{tree}=\sum_{k=1}^n\tr W_k(\Phi_k)+\sum_{k=1}^{n-1}\tr\lp \Phi_{k+1}\bar{X}_{k+1,k} X_{k,k+1}- \Phi_k X_{k,k+1}\bar{X}_{k+1,k} \rp
\ee

From these we easily obtain
\be\label{eqmotadj}
W_k^\prime(\Phi_k)+\bar{X}_{k,k-1}X_{k-1,k}-\X\Xb=0
\ee

\subsection{Coulomb vacua}

We first consider Coulomb vacua. We look for solutions of \eqref{eqmotbif} and \eqref{eqmotadj} with
\be\label{coulbif}
\X=\Xb=0\qquad\qquad 1\leqslant k\leqslant n-1
\ee

We choose a basis for the gauge group generators such that all $\Phi_k$'s are diagonal. Equations \eqref{eqmotadj} and \eqref{coulbif} thus impose that the eigenvalues of each $\Phi_k$ equal one of the roots of $W_k^\prime(z)=0$
\be
\Phi_k=\begin{pmatrix} a_{k,1}&0&\dots&\dots& &\dots &0\\0&a_{k,1}&0&\dots& &\dots&0\\ &\vdots &\ddots & & & &\\ & &\dots &a_{k,2}&\dots\\ & &  & &\ddots & & \vdots\\& & & & &\dots &a_{k,n_k}\end{pmatrix}
\ee
where each eigenvalue $a_{k,i}$ appears $N_{k,i}$ times (we may also have $N_{k,i}=0$ for some $i$).

Each factor of the gauge group is thus classically broken
\be
U(N_k)\quad\rightarrow\quad\prod_i U(N_{k,i})
\ee
where the product is taken over all non-zero $N_{k,i}$'s.

The $SU(N_{k,i})$ factors in $U(N_{k,i})$ actually confine due to quantum effects, and thus the quantum vacuum gauge group is broken to $U(1)^{m_k}$, where $m_k$ is the number of non-zero $N_{k,i}$'s.

\subsection{Higgs vacua}

We are now looking for vacua where some of the bifundamentals are not identically zero. We will make a simplifying assumption: only a pair \X, \Xb\, will be non-zero. This makes equations a little easier to be solved without spoiling more general Higgs vacua of any of their characteristics. We are also justified by the fact that this will actually be the case for an $A_2$ quiver, which is what we will eventually be interested in.

The equations of motion become in this case
\be\label{eqHiggs}
\begin{split}
&W_m^\prime(\Phi_m)=0\qquad\qquad\qquad m\neq k,k+1\\
&W_k^\prime(\Phi_k)-\X\Xb=0\\
&W_{k+1}^\prime(\Phi_{k+1})+\Xb\X=0
\end{split}
\ee
while \eqref{eqmotbif} and \eqref{combif} are still valid.

Eq. \eqref{combif} allows us to choose a basis such that $\Phi_i$, $X_{i,i+1}\bar{X}_{i+1,i}$ and $\bar{X}_{i,i+1}X_{i+1,i}$ are simultaneously diagonal for all $i$'s.

Let $\phi_{m,i}$ be the $i^{th}$ eigenvalue of $\Phi_m$, then the first of \eqref{eqHiggs} reads
\be
W_m^\prime(\phi_{m,i})=0\qquad\qquad\qquad m\neq k,k+1
\ee
which means that $\phi_{m,i}$ must be one of the roots $a_{m,j}$ of $W_m^\prime(z)=0$.

Now let $x_{i\j}$ be the elements of \X~ and $\bar{x}_{\i j}$ those of \Xb, with $i,j=1,\ldots,N_k$ and $\i,\j=1,\ldots,N_{k+1}$. Let also $s_1,\ldots,s_{N_k}$ and $t_1,\ldots,t_{N_{k+1}}$ be the diagonal entries of the fields \X\Xb~ and \Xb\X~ respectively. We have
\be \label{eqblock}
\begin{split}
&W_k^\prime(\phi_{k,i})-s_i=0\\
&W_{k+1}^\prime(\phi_{k+1,\j})+t_{\j}=0
\end{split}
\ee
and from \eqref{eqmotbif} (no sum over $\i$ or $j$ is understood)
\be
\begin{split}
&\bar{x}_{\i j}\phi_j=\phi_{\i} \bar{x}_{\i j}\\
&x_{j\i}\phi_{\i}=\phi_j x_{j \i}
\end{split}
\ee

This last equation tells us that $x_{i \j}$ and $\bar{x}_{\j i}$ can only be non-zero if $\phi_{k,i}$ is equal to $\phi_{k+1,\j}$. Equivalently if the $i^{th}$ eigenvalue of $\Phi_k$ is different from all eigenvalues of $\Phi_{k+1}$, then the $i^{th}$ row of \X~ and the $i^{th}$ column of \Xb~ are identically zero. The same is true for the $\j^{th}$ eigenvalue of $\Phi_{k+1}$ and the $\j^{th}$ row of \Xb~ and the $\j^{th}$ column of \X. We can thus choose a basis for the gauge group generators such that all common eigenvalues appear, with the same order, at the beginning of the diagonal in $\Phi_k$ and $\Phi_{k+1}$, and all non-common eigenvalues are placed at the end of the diagonal. Let $\varphi_i$ be the common eigenvalues, with multiplicity $r_i$ in $\Phi_k$ and $p_i$ in $\Phi_{k+1}$, and $\lambda_j$ and $\rho_l$ the non-common eigenvalues of $\Phi_k$ and  $\Phi_{k+1}$ with multiplicities $\alpha_j$ and $\beta_l$ respectively. We thus have that \X~ and \Xb~ are block-diagonal: in the $i^{th}$ block, \X~ has $r_i$ rows and $p_i$ columns, while \Xb~ has $p_i$ rows and $r_i$ columns. The last $\sum \alpha_j$ rows (columns) and $\sum \beta_k$ columns (rows) of \X (\Xb) are identically zero.

Equation \eqref{eqblock} and the block structure of \X ~and \Xb~ tell us that $\lambda_j=a_{k,m}$ for some $m=1,\ldots,n_k$ and $\rho_l=a_{k+1,p}$ for some $p=1,\ldots,n_{k+1}$.

Let $z_i=W_k^\prime(\varphi_i)$ and $\tilde{z}_j=W_{k+1}^\prime(\varphi_j)$, we then have from \eqref{eqblock} that $z_i$ is an eigenvalue of \X\Xb\, with multiplicity $r_i$ and $\tilde{z}_j$ is an eigenvalue of \Xb\X\, with multiplicity $p_i$. The requirements that  \X\Xb\, and \Xb\X\, be simultaneously diagonal, with the eigenvalue structure just mentioned, and that \X\, and \Xb\, be non-zero further impose that $r_i=p_i$. It follows then that $s_i=t_i$, and from \eqref{eqblock} that every $\varphi_i$ must be a solution of
\be\label{condhiggs}
W_k^\prime(z)+W_{k+1}^\prime(z)=0
\ee

\subsection{Classical phases of $A_2$}

We apply the results we have just found to the simplest non-affine quiver $A_2$\footnote{In this case, imposing that only one pair of bifundamentals is non-zero is the actual situation and not just a simplifying assumption.}.

We have\footnote{To fix notations we take $n_1>n_2$.}
\be
\begin{split}
&W_1^\prime(z)=g_{1,n_1+1}\prod_{i=1}^{n_1}(z-a_i)\\
&W_2^\prime(z)=g_{2,n_2+1}\prod_{j=1}^{n_2}(z-b_j)\\
&W_1^\prime(z)+W_2^\prime(z)=g_{1,n_1+1}\prod_{k=1}^{n_1}(z-c_k)
\end{split}
\ee

In the Coulomb phase the two bifundamentals $X\equiv X_{1,2}$ and $\bar{X}\equiv \bar{X}_{2,1}$ are classically zero, while the adjoint fields are given by
\be
\Phi_1=\begin{pmatrix} a_1&0&\dots&\dots& &\dots &0\\0&a_1&0&\dots& &\dots&0\\ &\vdots &\ddots & & & &\\ & &\dots &a_2&\dots\\ & &  & &\ddots & & \vdots\\& & & & &\dots &a_{n_1}\end{pmatrix}\qquad \Phi_2=\begin{pmatrix} b_1&0&\dots&\dots& &\dots &0\\0&b_1&0&\dots& &\dots&0\\ &\vdots &\ddots & & & &\\ & &\dots &b_2&\dots\\ & &  & &\ddots & & \vdots\\& & & & &\dots &b_{n_2}\end{pmatrix}
\ee
where each $a_i$ ($b_j$) appears $N_{1,i}$ ($N_{2,j}$) times, and some of the $N_{1,i}$ and $N_{2,j}$ can also be zero.

The $U(N_1)\times U(N_2)$ gauge group is broken to
\be\label{breakc}
\prod_i U(N_{1,i}) \times \prod_j U(N_{2,j})\quad\rightarrow\quad U(1)^{m_1+m_2}
\ee
where the products are taken on the $m_1$ and $m_2$ indices for which $N_{1,i}$ and $N_{2,j}$ are non-zero, and the second symmetry breaking is due to quantum effects which make the $SU(N)$ factors confining at low energies.

In the Higgs phase both adjoint and bifundamentals have non-zero classical VEV. We have
\be\label{higgsbifA2}
X=\begin{pmatrix} x_1&0&\dots&\dots& &\dots &0\\0&x_2&0&\dots& &\dots&0\\ \vdots &\vdots &\ddots & & & &\vdots\\ & &\dots &x_L&0&\dots&0\\& & \dots &0 &0 &\dots & 0\\ \vdots& & &\vdots &\vdots & &\vdots\\ \vdots& & &\vdots &\vdots & &\vdots \\0&\dots & & & &\dots &0\end{pmatrix}\qquad \bar{X}=\begin{pmatrix} \bar{x}_1&0&\dots&\dots& &\dots &\dots &0\\0&\bar{x}_2&0&\dots& &\dots&\dots &0\\ \vdots &\vdots &\ddots & & & & &\vdots\\ & &\dots &\bar{x}_L&0&\dots&\dots &0\\& & \dots &0 &0 &\dots & \dots& 0\\ \vdots& & &\vdots &\vdots & & &\vdots\\0&\dots & & & &\dots&\dots &0\end{pmatrix}
\ee
The adjoints are given by
\be
\Phi_1=\begin{pmatrix} c_1&0&\dots&\dots& &\dots &0\\0&\ddots&0&\dots& &\dots&0\\ \vdots&0 &c_{n_1} & 0&\ldots & &\\ & &\dots &a_1&\dots\\ & &  & &\ddots & & \vdots\\& & & & &\dots &a_{n_1}\end{pmatrix}\qquad \Phi_2=\begin{pmatrix} c_1&0&\dots&\dots& &\dots &0\\0&\ddots&0&\dots& &\dots&0\\ \vdots&0 &c_{n_1} & 0&\ldots & &\\ & &\dots &b_1&\dots\\ & &  & &\ddots & & \vdots\\& & & & &\dots &b_{n_1}\end{pmatrix}
\ee
where there are $L$ non-zero $c$ eigenvalues and each $c_k$ appears $r_k$ times and is placed in the same positions in both matrices. Each $a_i$ appears $N_{1,i}$ times in $\Phi_1$, each $b_j$ appears $N_{2,j}$ times in $\Phi_2$ and some of the $N_{1,i}$ and $N_{2,j}$ can be zero. The multiplicities must satisfy
\be
\begin{split}
&\sum_{i=1}^{n_1} N_{1,i}+\sum_{k=1}^{n_1}r_k=N_1\\
&\sum_{j=1}^{n_2} N_{2,j}+\sum_{k=1}^{n_1}r_k=N_2
\end{split}
\ee
The elements $x_1,\ldots ,x_L$ and $\bar{x}_1,\ldots ,\bar{x}_L$ in \eqref{higgsbifA2} are such that ($c_{k(i)}$ is the $i^{th}$ eigenvalue of either adjoints)
\be
x_i \bar{x}_i = W_1^\prime(c_{k(i)})=-W_2^\prime(c_{k(i)})
\ee

The $U(N_1)\times U(N_2)$ gauge group is broken to 
\be\label{breakh}
\prod_i U(N_{1,i}) \times \prod_j U(N_{2,j}) \prod_k U(r_k) \quad\rightarrow\quad U(1)^{m_1+m_2+L}
\ee

\section{Geometric data of $A_2$}\label{geom}

In this section we study the geometrical structure of the $A_2$ quiver. We start by analyzing the Riemann surface associated with the $\mathcal{N}=1$ theory, then write the anomaly equations that must be satisfied by the functions \eqref{operators} that describe the low-energy dynamics of the theory, and solve them for the one-form which as in \cite{CSW3} will be used to minimize the effective superpotential.

\subsection{3-fold Riemann surface $\Sigma$}

The algebraic curve associated with the $A_2$ quiver theory has been studied in a few different ways, via geometrical engineering in \cite{{CVq},{ohtatar}}, by means of matrix models \cite{lazaroiu}, and using generalized Konishi anomalies in \cite{{naculich},{quiver}}. The result is a non-hyperelliptic curve $\Sigma$ that can be written as
\be\label{curve}
F(y,z)\equiv y^3- p_2(z)y - p_3(z)=0
\ee
where
\be\label{p2p3}
\begin{split}
p_2(z) \equiv \,&t_{1}(z)^2 -  t_{1}(z)t_{2}(z)+t_{2}(z)^2 +\frac{1}{4} f_{1}(z) +\frac{1}{4}f_{2}(z)  \\
p_3(z) \equiv &-t_{1}(z)t_{2}(z)( t_{1}(z)- t_{2}(z) ) +\frac{1}{4} t_{1}(z)f_{2}(z) 
-\frac{1}{4} t_{2}(z)f_{1}(z)+\\&+
\frac{1}{4}g_{1}(z) +\frac{1}{4} g_{2}(z)
\end{split}
\ee
and
\be\label{t1t2}
\begin{split}
t_{1}(z) &\equiv \frac{2}{3} W'_{1}(z) + \frac{1}{3} W'_{2}(z) \\
t_{2}(z) &\equiv \frac{1}{3} W'_{1}(z) + \frac{2}{3} W'_{2}(z)
\end{split}
\ee
while $f_1(z)$, $g_1(z)$ are polynomials of degree $n_1-1$, and $f_2(z)$, $g_2(z)$ are polynomials of degree $n_2-1$.

Let $y_1(z)$, $y_2(z)$ and $y_3(z)$ be the three complex roots of \eqref{curve}, then since there is no term quadratic in $y$, they sum up to zero. We have then $y_3(z)=-y_1(z)-y_2(z)$. Matrix model loop equations \cite{lazaroiu} and Ward identities for generalized anomalous Konishi transformations \cite{{naculich},{quiver}} allow to express the two independent roots in terms of the resolvents $R_1(z)$ and $R_2(z)$ of the theory
\be\label{roots}
\begin{split}
y_{1}(z)&= R_{1}(z) - \frac{2}{3} W'_{1}(z) - \frac{1}{3} W'_{2}(z) \\
y_{2}(z)&= -R_{2}(z) + \frac{1}{3} W'_{1}(z) + \frac{2}{3} W'_{2}(z)
\end{split}
\ee

Being written as a cubic complex equation, the curve $\Sigma$ \eqref{curve}, is a 3-fold non-hyperelliptic Riemann surface, each sheet being described by one of the roots $y_i(z)$. There are some special points on $\Sigma$ where two separated sheets meet. The Dini or implicit function theorem tells us how to find these branch points. They are solutions of the system of two equations
\be\label{implicit}
\begin{split}
&F(z,y)=0\\
&\frac{\partial F}{\partial y}(y,z)=0
\end{split}
\ee

The conditions that must be satisfied by the branch points read
\be\label{branchgen}
\begin{split}
&y=-\frac{3}{2}\frac{p_3(z)}{p_2(z)}\\
&p_3^2(z)-\frac{4}{27}p_2^3(z)=0
\end{split}
\ee

As a first step we consider the classical limit of the curve. This limit corresponds to taking $f_1(z)=f_2(z)=g_1(z)=g_2(z)=0$, since in the field theory derivation of \eqref{curve} it is shown that they are given by the expectation values of operators containing a bilinear in the superfield strength of the gauge symmetry \cite{{naculich},{quiver}}. Using equations \eqref{p2p3} and \eqref{t1t2} in \eqref{branchgen} we find that the branch points are located at
\be\label{branchclass}
\begin{split}
&W_1^\prime(z)^2 ~W_2^\prime(z)^2~\lp W_1^\prime(z)+ W_2^\prime(z) \rp^2=0\\
&y=\frac{\lp W_1^\prime(z)- W_2^\prime(z) \rp\lp 2W_1^\prime(z)+ W_2^\prime(z) \rp\lp W_1^\prime(z)+2 W_2^\prime(z) \rp}{6\lp W_1^\prime(z)^2+W_1^\prime(z)W_2^\prime(z)+W_2^\prime(z)^2\rp^3}
\end{split}
\ee

Let us classify the solutions according to which factor of the first equation is zero
\be\label{abc}
\begin{split}
&W_1^\prime(a_i)=0\\
&W_2^\prime(b_j)=0\\
&W_1^\prime(c_k)+W_2^\prime(c_k)=0
\end{split}
\ee
where $i=1,\ldots,n_1$, $j=1,\ldots,n_2$ and $k=1, \ldots, n_1$. The $z$ coordinate of the branch points satisfies the same conditions as the classically allowed eigenvalues of the adjoint fields $\Phi_1$ and $\Phi_2$ in the Coulomb and Higgs phases.

The first equation in \eqref{branchclass} tells us that classically all the branch points are actually double roots. Beyond the classical approximation, these double roots will in general open up into cuts which in a small coupling regime are still localized near the classical branch points. The splitting is governed by the polynomials $f_1(z)$, $f_2(z)$ and $g(z)\equiv g_1(z)+g_2(z)$, which through \eqref{fielddegrees} are related to the expectation values of the superfield strength bilinears. In the quantum theory the three sheets are smoothly connected through the cuts, and actually when we stay close to them there is no clear separation between two sheets.

Let us now determine how the sheets are connected to each other through the branch points. From \eqref{roots} we see that (since we are considering the classical limit, $R_1(z)=R_2(z)=0$) we have 
\be
\begin{split}
&W_1^\prime (a_i)=0\quad\rightarrow\quad \left\{ \begin{array}{l} y_1(a_i)=y_3(a_i)=-\frac{1}{3}W_2^\prime(a_i) \\ y_2(a_i)=\frac{2}{3}W_2^\prime(a_i)\end{array} \right.\\
&W_2^\prime (b_j)=0\quad\rightarrow\quad \left\{ \begin{array}{l} y_2(b_j)=y_3(b_j)=\frac{1}{3}W_1^\prime(b_j) \\ y_1(b_j)=-\frac{2}{3}W_1^\prime(b_j)\end{array} \right.\\
&W_1^\prime (c_k)+ W_2^\prime (c_k)=0\quad\rightarrow\quad \left\{ \begin{array}{l} y_1(c_k)=y_2(c_k)=-\frac{1}{3}W_1^\prime(c_k) \\ y_3(c_k)=\frac{2}{3}W_1^\prime(c_k)\end{array} \right.
\end{split}
\ee

The third sheet is connected to the others only through Coulomb branch points, while the first and second sheets communicate through Higgs branch points, as shown in Figure \ref{figcuts}.
\FIGURE[!ht]{\epsfig{file=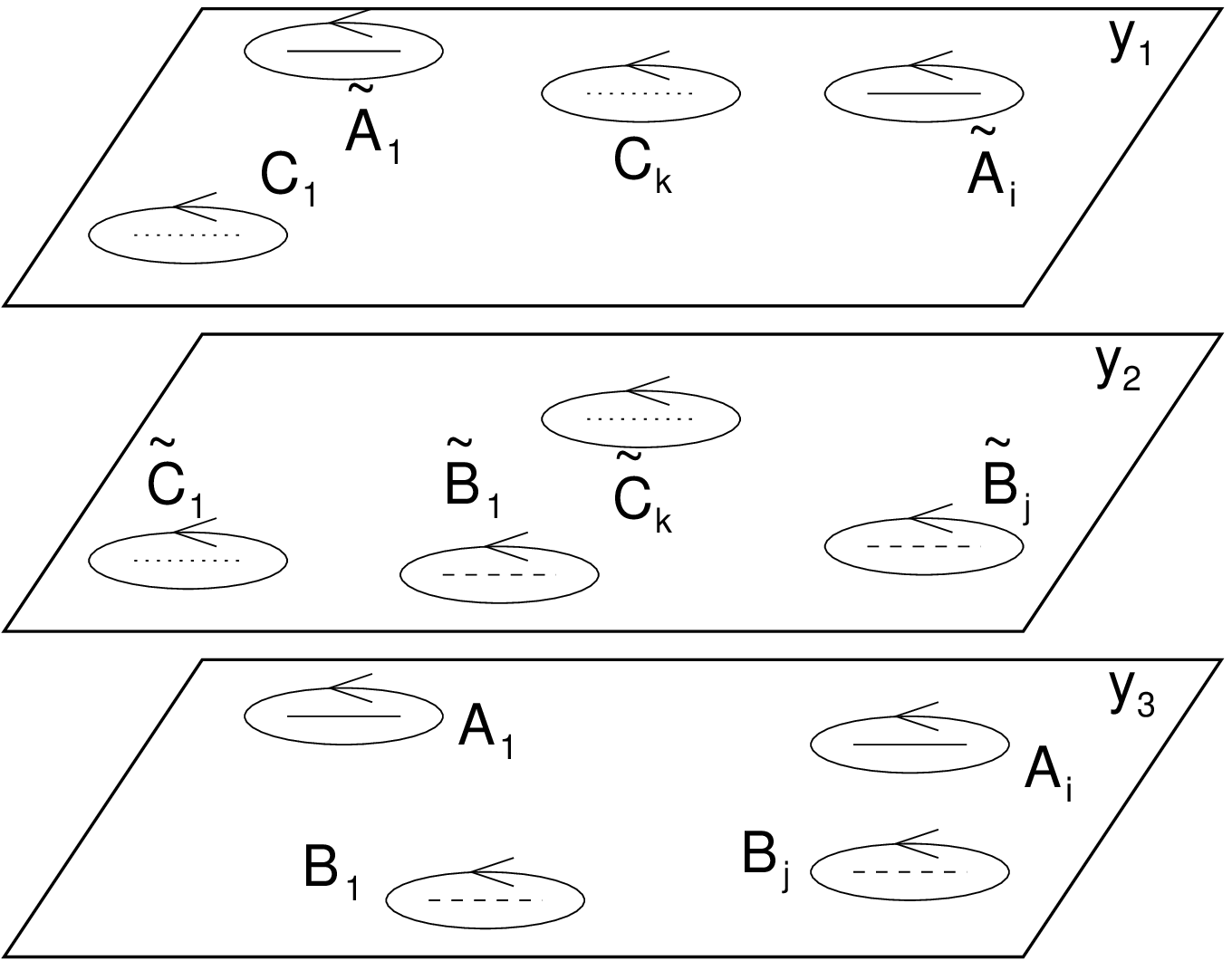,width=0.8\textwidth}\caption{The three-sheet structure of the Riemann surface $\Sigma$ is shown here. When quantum corrections to the geometry of the surface are taken into account, classical branch points connecting two different sheets blow up into cuts. Cuts $A_i$ and $B_j$ are associated with Coulomb branch points, while cuts $C_k$ come from Higgs branch points.}\label{figcuts}}

\subsection{Anomaly equations}

In \cite{{naculich},{quiver}} a set of two equations that must be satisfied by the resolvents of the gauge theory was obtained. The resolvent is not the only interesting operator that we need to describe the low-energy dynamics of a gauge theory. The dynamics is determined also by the other two operators appearing in \eqref{operators} \cite{CSW1}. We thus need a set of equations that these quantities must satisfy. In \cite{CSW1} a simple method was shown to derive such equations. We add to the complex plane a fermionic coordinate $\psi_\alpha$, and define functions of the coordinate $z$ as coefficients of the power expansion in $\psi_\alpha$ of some function defined on the enlarged space. In particular we take
\be\label{superresolvents}
\begin{split}
&\mathcal{R}_1(z,\psi_\alpha)\equiv-\frac{1}{2}\;\tr\!\!\lp\frac{1}{4\pi}\mathcal{W}_{1\alpha}-\psi_\alpha\rp^2\frac{1}{z-\Phi_1}=R_1(z)+\psi_\alpha w_1^\alpha(z)-\psi^1\psi^2 T_1(z)\\
&\mathcal{R}_2(z,\psi_\alpha)\equiv-\frac{1}{2}\;\tr\!\!\lp\frac{1}{4\pi}\mathcal{W}_{2\alpha}-\psi_\alpha\rp^2\frac{1}{z-\Phi_2}=R_2(z)+\psi_\alpha w_2^\alpha(z)-\psi^1\psi^2 T_2(z)
\end{split}
\ee

Since Ward identities do not contain any explicit dependence on the fermionic coordinate $\psi_\alpha$ \cite{CSW1}, the equations satisfied by these pseudo-superoperators are the same as those satisfied by their lowest order components, that is by $R_1(z)$ and $R_2(z)$ \cite{{naculich},{quiver}}
\be\label{superquad}
\begin{split}
&W_1^\prime(z)\mathcal{R}_1(z,\psi_\alpha)+\frac{1}{4}\mathfrak{f}_1(z,\psi_\alpha)+W_2^\prime(z)\mathcal{R}_2(z,\psi_\alpha)+\frac{1}{4}\mathfrak{f}_2(z,\psi_\alpha)+\mathcal{R}_1(z,\psi_\alpha)\mathcal{R}_2(z,\psi_\alpha)=\\&\qquad\qquad=\mathcal{R}_1(z,\psi_\alpha)^2+\mathcal{R}_2(z,\psi_\alpha)^2
\end{split}
\ee
\be\label{supercub}
\begin{split}
&\mathcal{R}_1(z,\psi_\alpha)^2\mathcal{R}_2(z,\psi_\alpha)-\mathcal{R}_1(z,\psi_\alpha)\mathcal{R}_2(z,\psi_\alpha)^2=W_1^\prime(z)\big( \mathcal{R}_1(z,\psi_\alpha)^2- W_1^\prime(z)\mathcal{R}_1(z,\psi_\alpha)+\\&\qquad\qquad-\frac{1}{4}\mathfrak{f}_1(z,\psi_\alpha)\big)-W_2^\prime(z)\big( \mathcal{R}_2(z,\psi_\alpha)^2- W_2^\prime(z)\mathcal{R}_2(z,\psi_\alpha)-\frac{1}{4}\mathfrak{f}_2(z,\psi_\alpha)\big)+\\&\qquad\qquad+\frac{1}{4}\mathfrak{g}_1(z,\psi_\alpha)+\frac{1}{4}\mathfrak{g}_2(z,\psi_\alpha)
\end{split}
\ee
where ($a=1,2$)
\be\label{polynomials}
\begin{split}
&\mathfrak{f}_a(z,\psi_\alpha)=-2\;\tr\!\!\lp \frac{1}{4\pi}\mathcal{W}_{a,\alpha}-\psi_\alpha\rp ^2  \frac{W_a^\prime(\Phi_a)-W_a^\prime(z)}{z-\Phi_a} =f_a(z) +\psi_\alpha\rho_a^\alpha(z)-\psi_1\psi_2l_a(z)\\
&\mathfrak{g}_a(z,\psi_\alpha)=-2\;\tr\!\!\lp\lp \frac{1}{4\pi}\mathcal{W}_{a,\alpha}-\psi_\alpha\rp ^2\frac{W_a^\prime(\Phi_a)-W_a^\prime(z)}{z-\Phi_a} ``XX\textrm{''}\rp =\\&\qquad\qquad=g_a(z) +\psi_\alpha\sigma_a^\alpha(z)-\psi_1\psi_2r_a(z)
\end{split}
\ee
and $``XX\textrm{''} $ stands for either $X_{12}\bar{X}_{21}$ or $\bar{X}_{21}X_{12}$, the choice being fixed by gauge invariance of the trace. All the polynomials labeled by $a=1$ are thus of degree $n_1-1$ while all polynomials with $a=2$ have degree $n_2-1$.

Expanding equations \eqref{superquad} and \eqref{supercub} in a power series in $\psi_\alpha$ will give three pairs of equations for the operators $R_a(z)$, $w_a^\alpha(z)$ and $T_a(z)$. We obtain three quadratic equations
\be\label{anomq}
\begin{split}
&W_1^\prime(z)R_1(z)+\frac{1}{4}f_1(z)+W_2^\prime(z)R_2(z)+\frac{1}{4}f_2(z)+R_1(z)R_2(z)=R_1(z)^2+R_2(z)^2\\
&W_1^\prime(z)w_1^\alpha(z)+\frac{1}{4}\rho_1^\alpha(z)+W_2^\prime(z)w_2^\alpha(z)+\frac{1}{4}\rho_2^\alpha(z)+R_1(z)w_2^\alpha(z)+R_2(z)w_1^\alpha(z)=\\&\qquad\quad=2R_1(z)w_1^\alpha(z)+2R_2(z)w_2^\alpha(z)\\
&W_1^\prime(z)T_1(z)+\frac{1}{4}l_1(z)+W_2^\prime(z)T_2(z)+\frac{1}{4}l_2(z)+R_1(z)T_2(z)+R_2(z)T_1(z)+\\&\qquad\quad +w_{1\alpha}(z)w_2^\alpha(z)=2R_1(z)T_1(z)+2R_2(z)T_2(z)+w_{1\alpha}(z)w_1^\alpha(z)+w_{2\alpha}(z)w_2^\alpha(z)
\end{split}
\ee
and three cubic equations
\be\label{anomc}
\begin{split}
&R_1(z)^2R_2(z)-R_1(z)R_2(z)^2=W_1^\prime(z)\big( R_1(z)^2-W_1^\prime(z)R_1(z)-\frac{1}{4}f_1(z)\big)+\\&\qquad\quad-W_2^\prime(z)\big( R_2(z)^2-W_2^\prime(z)R_2(z)-\frac{1}{4}f_2(z)\big) +\frac{1}{4}g_1(z)+\frac{1}{4}g_2(z)\\
&R_1(z)^2w_2^\alpha(z)-R_2(z)^2w_1^\alpha(z)+2R_1(z)R_2(z)\big(w_1^\alpha(z)-w_2^\alpha(z)\big)=\!W_1^\prime(z)\big(2R_1(z)w_1^\alpha(z)+\\&\qquad\quad-W_1^\prime(z)w_1^\alpha(z)-\frac{1}{4}\rho_1^\alpha(z)\big)-W_2^\prime(z)\big(2R_2(z)w_2^\alpha(z)-W_2^\prime(z)w_2^\alpha(z)-\frac{1}{4}\rho_2^\alpha(z)\big)+\\&\qquad\quad+\frac{1}{4}\sigma_1^\alpha(z)+\frac{1}{4}\sigma_2^\alpha(z)
\end{split}
\ee
\be\nonumber
\begin{split}
&R_1(z)^2T_2(z)-R_2(z)^2T_1(z)+2R_1(z)R_2(z)\big(T_1(z)-T_2(z)\big)-R_1(z)w_{2\alpha}(z)w_2^\alpha(z)+\\&\qquad\quad+R_2(z)w_{1\alpha}(z)w_1^\alpha(z)+2\big(R_1(z)-R_2(z)\big)w_{1\alpha}(z)w_2^\alpha(z)=\frac{1}{4}r_1(z)+\frac{1}{4}r_2(z)+\\&\qquad\quad+W_1^\prime(z)\big( 2R_1(z)T_1(z)-W_1^\prime(z)T_1(z)+w_{1\alpha}(z)w_1^\alpha(z)-\frac{1}{4}l_1(z) \big)+\\&\qquad\quad-W_2^\prime(z)\big( 2R_2(z)T_2(z)-W_2^\prime(z)T_2(z)+w_{2\alpha}(z)w_2^\alpha(z)-\frac{1}{4}l_2(z) \big)
\end{split}
\ee

\subsection{A meromorphic one-form $\sigma(z)dz$ on $\Sigma$}

We look for a supersymmetric solution to equations \eqref{anomq} and \eqref{anomc}. Being vacuum expectation values of fermionic operators, $w_1^\alpha(z)$ and $w_2^\alpha(z)$ will both vanish in a supersymmetric vacuum. Since we have already given a solution for the resolvents as roots \eqref{roots} of a complex cubic equation, we only have to find $T_1(z)$ and $T_2(z)$. We will show that these two functions will be proportional to the coefficients, on different sheets of $\Sigma$, of a meromorphic one-form $\sigma(z)dz$ defined on the whole surface $\Sigma$.

Substituting $w_a^\alpha(z)=0$ into \eqref{anomq} and\eqref{anomc}, and making use of \eqref{roots} and a little algebra we find
\be\label{T1T2}
\begin{split}
T_1(z)=&-\frac{ y_1(z)\lp l_1(z)+l_2(z)\rp} {4\lp y_1(z)-y_2(z)\rp\lp 2 y_1(z)+y_2(z)\rp}+\\&-\frac{\lp 5l_1(z)+2l_2(z)\rp W_1^\prime(z)-\lp 2l_1(z)+5l_2(z)\rp W_2^\prime(z)-3\lp r_1(z)+r_2(z)\rp}{12\lp y_1(z)-y_2(z)\rp\lp 2 y_1(z)+y_2(z)\rp}\\ \\
T_2(z)=&-\frac{ y_2(z)\lp l_1(z)+l_2(z)\rp} {4\lp y_1(z)-y_2(z)\rp\lp  y_1(z)+2 y_2(z)\rp}+\\&-\frac{\lp 5l_1(z)+2l_2(z)\rp W_1^\prime(z)-\lp 2l_1(z)+5l_2(z)\rp W_2^\prime(z)-3\lp r_1(z)+r_2(z)\rp}{12\lp y_1(z)-y_2(z)\rp\lp  y_1(z)+2 y_2(z)\rp}
\end{split}
\ee

We notice that $T_1(z)$ is singular when $y_1(z)=y_2(z)$ and when $y_1(z)=y_3(z)$, while $T_2(z)$ is singular when $y_1(z)=y_2(z)$ and when $y_2(z)=y_3(z)$. It is natural then to look at $T_1$ and $T_2$ as functions defined only on the first and second sheet of $\Sigma$ respectively.

There is something more to this. Let us consider the derivative with respect to $y$ of the function $F(y,z)$ that defines $\Sigma$ through \eqref{curve}. On the first sheet we have\footnote{We use the relation $y_1(z)^2+y_2(z)^2+y_1(z)y_2(z)=p_2(z)$ which follows from $y_1(z)$, $y_2(z)$ and $y_3(z)=-y_1(z)-y_2(z)$ being roots of \eqref{curve}.}
\be
\frac{\partial F}{\partial y}(y_1(z),z)=\lp y_1(z)-y_2(z)\rp \lp 2y_1(z)+y_2(z)\rp
\ee
and on the second one
\be
\frac{\partial F}{\partial y}(y_2(z),z)=-\lp y_1(z)-y_2(z)\rp \lp y_1(z)+2y_2(z)\rp
\ee
which, apart from a sign and multiplicative constants, match the denominators in \eqref{T1T2}. This suggests to define a $1$-form on $\Sigma$
\be\label{sigmaform}
\begin{split}
\sigma(z)dz=&-\frac{1}{4}\lp l_1(z)+l_2(z)\rp\frac{ y(z)} {\frac{\partial F}{\partial y}(y(z),z)}dz+\\&-\frac{1}{12}\frac{\lp 5l_1(z)+2l_2(z)\rp W_1^\prime(z)-\lp 2l_1(z)+5l_2(z)\rp W_2^\prime(z)\!-3\lp r_1(z)+r_2(z)\rp}{\frac{\partial F}{\partial y}(y(z),z)}dz
\end{split}
\ee
which on the first sheet is equal to $T_1(z)dz$ and on the second sheet is $-T_2(z)dz$.

Equations \eqref{polynomials} tell us that $l_1(z)$ and $r_1(z)$ are degree $n_1-1$ polynomials, while $l_2(z)$ and $r_2(z)$ are of degree $n_2-1$, thus $l_1(z)+l_2(z)$ and $r_1(z)+r_2(z)$ are of degree $n_1-1$, and the numerator of the second term in \eqref{sigmaform} has degree $2 n_1-1$. Because of the results of Appendix \ref{holforms}, this makes of  $\sigma(z)dz$ a meromorphic one-form on $\Sigma$, regular at the branch points, but singular at infinity. 

There are $2n_1+n_2$ free parameters to fix in $\sigma(z)dz$, the coefficients of $l_1(z)$, $l_2(z)$ and $r(z)\equiv r_1(z)+r_2(z)$\footnote{Since in all physically relevant formulas only the sum of $r_1(z)$ and $r_2(z)$ appears, they contribute only $n_1$ parameters.}. They can be determined by imposing conditions on the periods of $\sigma(z)dz$ around the cycles $A_i$, $B_j$ and $C_k$. There are exactly $2 n_1+n_2$ such cycles. We impose then
\be\label{condsigma}
\oint_{A_i} dz ~\sigma(z) =-N_{1,i} \qquad \oint_{B_j} dz ~\sigma(z) =N_{2,j} \qquad \oint_{C_k} dz ~\sigma(z) =r_k
\ee
where classically $N_{1,i}$ ($N_{2,j}$) is the number of eigenvalues of $\Phi_1$ ($\Phi_2$) which equal $a_i$ ($b_j$), and $r_k$ is the number of eigenvalues in $\Phi_1$ and $\Phi_2$ that are equal to $c_k$.

Let $P_1$, $P_2$ and $P_3$ be the points at infinity on the three sheets respectively. Then $\sigma(z)dz$ has simple poles at infinity, with residues $-N_1$ in $P_1$, $N_2$ in $P_2$ and $N_1-N_2$ in $P_3$.

\section{Matrix model computation of the effective superpotential}\label{matrixpot}

In this section we study the effective superpotential of the theory using the matrix model approach.

We use the prescription of \cite{DV}: the fields of the gauge theory are translated into matrices
\be
\begin{split}
\Phi_k\qquad&\longrightarrow\qquad\pk\\
\X\qquad&\longrightarrow\qquad\xk\\
\Xb\qquad&\longrightarrow\qquad\xbk
\end{split}
\ee
whose ranks are taken to be very large in all computations. In this limit the model is governed by a set of t'Hooft coupling constants 
\be
\s=g_{\scriptscriptstyle{S}} \hat{N}_k
\ee
where $g_{\scriptscriptstyle{S}}$ is the coupling constant of the underlying open string theory and $\hat{N}_k$ is the rank of the matrix $\hat{\Phi}_k$, which has no connection with the rank $N_k$ of the gauge group $U(N_k)$.

The potential of the matrix model is the same as the superpotential of the gauge theory with superfields substituted by the corresponding matrices
\be
W_{tree}=\sum_{k=1}^n\tr W_k(\pk) +\sum_{k=1}^{n-1}\tr\lp\xbk\pk\xk-\xk\hat{\Phi}_{k+1}\xbk\rp
\ee

\subsection{Sphere and disk contributions to the free energy}

First of all we want to evaluate $\mathcal{F}_0$ and $\mathcal{F}_1$ for a $A_n$ quiver
\be\label{zmat}
e^{-\gq\mathcal{F}_0-\g\mathcal{F}_1+\ldots}=\frac{1}{\prod_{k=1}^{n}\mathrm{Vol}~U(\hat{N}_k)}\int\lp\prod_{k=1}^n\frac{d\pk}{\Lambda_k^{\hat{N}_k^2}}\rp\lp\prod_{k=1}^{n-1}\frac{d\xk d\xbk}{\nu_k^{\hat{N}_k\hat{N}_{k+1}}}\rp e^{-\g W_{tree}}
\ee
where $\Lambda_k$ is the dimension one scale of the gauge theory factor $U(N_k)$, and $\nu_k$ is a dimensionless normalization constant which we will later fix.

The first step is to integrate out the bifundamental matrices. This is possible when all the \pk's are diagonal. Letting $\lambda_{k,i}$ be the $i^{th}$ eigenvalue of \pk, we find
\be\label{outbif}
\begin{split}
\int d\xk d\xbk e^{-\g\tr\lp\xbk\pk\xk-\xk\hat{\Phi}_{k+1}\xbk\rp} &= e^{ -\lp\hat{N}_k\hat{N}_{k+1}\ln \frac{1}{2\pi g_{\scriptscriptstyle{S}}}+ \sum_{i=1}^{\hat{N}_k}\sum_{j=1}^{\hat{N}_{k+1}}\ln(\lambda_{k,i}-\lambda_{k+1,j})\rp}=\\
&=e^{ -\gq \hat{S}_k\hat{S}_{k+1}\ln \frac{1}{2\pi g_{\scriptscriptstyle{S}}}} ~e^{-\sum_{i=1}^{\hat{N}_k}\sum_{j=1}^{\hat{N}_{k+1}}\ln(\lambda_{k,i}-\lambda_{k+1,j})}
\end{split}
\ee
where the first factor in the second line is independent of the eigenvalues $\lambda_{k,i}$, and for convenience it will be written as a constant $B$ in some of the following formulas.

We now substitute \eqref{outbif} into \eqref{zmat} and evaluate the adjoint matrices integrals as integrals over their eigenvalues $\lambda_{k,i}$. In doing so we have to add the Vandermonde term inside the integral. Neglecting, only for the moment, the $\Lambda_k$ and $\nu_k$ dependences, we find for the right-hand side of \eqref{zmat}
\be\label{eigenint}
B\int( \prod_{k=1}^n \prod_{i=1}^{N_k}d\lambda_{k,i}) e^{\sum_{k=1}^n\lp -\g\sum_i W_k(\lambda_{k,i})+2
\sum_{i<j}\ln(\lambda_{k,i}-\lambda_{k,j})\rp-\sum_{k=1}^{n-1}\sum_{i=1}^{N_k}\sum_{j=1}^{N_{k+1}}\ln(\lambda_{k,i}-\lambda_{k+1,j}) }
\ee

The interaction with the bifundamentals and quantum effects (which are encoded in the Vandermonde determinant) introduce an effective action for the eigenvalues of the adjoint matrices, from which we can write the quantum equations of motion for a generic probe eigenvalue $x\equiv\lambda_{k,N_k+1}$\footnote{This equation and the following \eqref{eqres} are valid only for $k\neq 1,n$. In these cases  either of the last two terms is not defined and is absent from the corresponding equation of motion.}
\be\label{effeqmot}
\g W_k^\prime (x)-\sum_{j=1}^{\hat{N}_k}\frac{2}{x-\lambda_{k,j}}+\sum_{j=1}^{\hat{N}_{k-1}}\frac{1}{x-\lambda_{k-1,j}}+\sum_{j=1}^{\hat{N}_{k+1}}\frac{1}{x-\lambda_{k+1,j}}=0
\ee 

The eigenvalues will  move off their classical value and spread over intervals which are in general still centered around the critical points of $W_k$. In the large $\hat{N}_k$ limit the distribution becomes continuous and is described by a set of distribution functions $\rho_k(x)$ which we take to be normalized as
\be\label{dens}
\int \mathrm{d}x ~\rho_k(x)=g_{\scriptscriptstyle S} \hat{N}_k=\hat{S}_k
\ee

This is the matrix model description of the splitting of coincident branch points into cuts connecting different sheets of $\Sigma$. Quantum corrections and bifundamental interactions give rise to Coulomb-like interactions between the eigenvalues of the adjoint matrices. 

If we define a set of matrix model resolvents
\be
\omega_k(x)=\frac{1}{\hat{N}_k}\sum_{i=1}^{N_k}\frac{1}{x-\lambda_{k,i}}
\ee
equation \eqref{effeqmot} becomes
\be\label{eqres}
W_k^\prime (x)-2\hat{S}_k\omega_k(x)+\hat{S}_{k-1}\omega_{k-1}(x)+\hat{S}_{k+1}\omega_{k+1}(x)=0
\ee
and the eigenvalue distribution is given by
\be\label{dens-res}
\rho_k(x)=-\frac{1}{2\pi i}~\mathrm{disc}~ \omega_k(z)\arrowvert_{z=x}
\ee

Let us now go on with the evaluation of the free energy of the matrix model. Integral \eqref{eigenint} can be evaluated via the saddle point approximation, and gives (we insert also the $\Lambda_k$ and $\nu_k$ contributions we had neglected)
\be\label{saddle}
\begin{split}
\exp&\left\{ -\gq \lp \sum_{k=1}^{n} \lp \int \mathrm{d}x \rho_k(x) W_k(x) -\int \mathrm{d}x ~\mathrm{d}x^\prime \rho_k(x)\rho_k(x^\prime) \ln \frac{\arrowvert x-x^\prime \arrowvert}{\Lambda_k}  \rp + \right.\right.\\ &\qquad\quad\left. \left.+\sum_{k=1}^{n-1}   \int \mathrm{d}x ~\mathrm{d}x^\prime \rho_k(x)\rho_{k+1}(x^\prime) \ln  \frac{\arrowvert x-x^\prime \arrowvert}{2\pi g_{\scriptscriptstyle S} \nu_k^{-1}  }      \rp \right\}
\end{split}
\ee

We can now fix $\nu_k$ to cancel the $g_{\scriptscriptstyle S}$ dependence in the argument of the last logarithm. This choice is suggested by the requirement that the free energy of the matrix model (and thus the low-energy superpotential of the gauge theory) brings no explicit memory of the underlying string theory. The only gauge theory dimensionful parameters are the dynamical scales $\Lambda_k$. We put then
\be
\nu_k=\frac{2\pi g_{\scriptscriptstyle S}}{\sqrt{\Lambda_k \Lambda_{k+1}}}
\ee

Comparing \eqref{saddle} with the left-hand side of \eqref{zmat}, we find
\be\label{freeen}
\begin{split}
\mathcal{F}_0=& \sum_{k=1}^{n} \lp \int \mathrm{d}x \rho_k(x) W_k(x) -\int \mathrm{d}x ~\mathrm{d}x^\prime \rho_k(x)\rho_k(x^\prime) \ln \frac{\arrowvert x-x^\prime \arrowvert}{\Lambda_k}  \rp +\\&+\sum_{k=1}^{n-1}   \int \mathrm{d}x ~\mathrm{d}x^\prime \rho_k(x)\rho_{k+1}(x^\prime) \ln \frac{\arrowvert x-x^\prime \arrowvert}{\sqrt{\Lambda_k \Lambda_{k+1}}} \\
\mathcal{F}_1=&~0
\end{split}
\ee

The disk contribution to the free energy vanishes as it had to be expected, since there is no fundamental matter in our theory.

\subsection{Effective superpotential}

The degrees of freedom of the matrix model are given by the periods of the resolvents around the compact cycles of $\Sigma$
\be\label{degrees}
\begin{split}
&\hat{S}_{1,i}=\frac{1}{2\pi i}\oint_{\tilde{A}_i} dz~\omega_1(z)\\
&\hat{S}_{2,j}=\frac{1}{2\pi i}\oint_{\tilde{B}_j} dz~\omega_2(z)\\
&\hat{H}_{1,k}=\frac{1}{2\pi i}\oint_{C_k} dz~\omega_1(z)\\
&\hat{H}_{2,k}=\frac{1}{2\pi i}\oint_{\tilde{C}_k} dz~\omega_2(z)\\
\end{split}
\ee
which satisfy the constraints (see \eqref{dens})
\be
\begin{split}
&\hat{S}_1=\sum_{i=1}^{n_1}\hat{S}_{1,i}+\sum_{k=1}^{n_1}\hat{H}_{1,k}\\
&\hat{S}_2=\sum_{j=1}^{n_2}\hat{S}_{2,j}+\sum_{k=1}^{n_1}\hat{H}_{2,k}
\end{split}
\ee

Because of \eqref{res/sheets} (which is the translation in the matrix model language of the gauge theory relation \eqref{roots} \cite{quiver}), $\omega_1(z)=-\omega_2(z)$ on the cuts $(c_k^-,c_k^+)$, thus
\be\label{H1=H2}
\hat{H}_{1,k}=\hat{H}_{2,k}\equiv \hat{H}_k
\ee
and they are, indeed,  the same period of $\Sigma$.

In the matrix model approach, the effective superpotential of the associated gauge theory can be written as a functional depending only on the geometric properties of the algebraic curve singled out by the resolvents of the matrix model \cite{{DV},{CSW1},{CSW3},{CSW2}}. This interpretation is easily generalizable to a wide class of theories, among which the $A_2$ quiver we are considering. We obtain, then, that the effective superpotential for such a gauge theory is given by
\be\label{We}
\begin{split}
W_{eff}=&\sum_{i=1}^{n_1}N_{1,i}\frac{\partial \mathcal{F}_0}{\partial \hat{S}_{1,i}}+\sum_{j=1}^{n_2}N_{2,j}\frac{\partial \mathcal{F}_0}{\partial \hat{S}_{2,j}}+\sum_{k=1}^{n_1}r_k\frac{\partial \mathcal{F}_0}{\partial \hat{H}_{k}}+\\&+2\pi i \lp\sum_{i=1}^{n_1-1}b_{1,i} \hat{S}_{1,i}+\sum_{j=1}^{n_2-1}b_{2,j} \hat{S}_{2,j}+\sum_{k=1}^{n_1}h_{k} \hat{H}_{k}\rp
\end{split}\ee
where $b_{1,i}$, $b_{2,j}$ and $h_k$ are integers selecting the vacuum we chose \cite{CSW2}.

Substituting in \eqref{We} the results we found in Appendix \ref{appmatcomp} for the derivatives of the free energy $\mathcal{F}_0$, and making use of \eqref{H1=H2} and the dictionary between matrix model and field theory degrees of freedom, the $A_2$ quiver gauge theory effective superpotential may be written as
\be\label{Weff}
\begin{split}
W_{eff}=&N_1 W_1(\Lambda_0)+N_2 W_2(\Lambda_0)-S_1\,\ln \!\lp\frac{(-\Lambda_0)^{2N_1-N_2}}{\Lambda_1^{2N_1-N_2}}\lp\frac{\Lambda_2}{\Lambda_1} \rp^{\frac{N_2}{2}}\rp+\\&-S_2\,\ln \!\lp\frac{(-\Lambda_0)^{2N_2-N_1}}{\Lambda_2^{2N_2-N_1}}\lp\frac{\Lambda_1}{\Lambda_2} \rp^{\frac{N_1}{2}}\rp-H\,\ln \!\lp\frac{(-\Lambda_0)^{N_1+N_2}}{\Lambda_1^{2N_1-N_2}\Lambda_2^{2N_2-N_1}}\lp\frac{\Lambda_2}{\Lambda_1} \rp^{\frac{N_1-N_2}{2}}\rp +\\&+\sum_{i=1}^{n_1}N_{1,i}\int_{B_{31}^i} dz~y(z)+\sum_{j=1}^{n_2}N_{2,j}\int_{B_{23}^j} dz~y(z)-\sum_{k=1}^{n_1}r_k\int_{B_{12}^k} dz~y(z)+\\&+2\pi i \lp\sum_{i=1}^{n_1-1}b_{1,i} S_{1,i}+\sum_{j=1}^{n_2-1}b_{2,j} S_{2,j}+\sum_{k=1}^{n_1}h_k H_{k}\rp
\end{split}
\ee
where  $S_{1,i}$, $S_{2,j}$ and $H_k$ were defined in \eqref{fielddegrees} and the paths $B_{31}^i$, $B_{23}^{j}$ and $B_{12}^k$ are  the paths connecting $P^3$ with $P^1$, $P^2$ with $P^3$, $P^1$ with  $P^2$ ($P_1$, $P_2$, and $P_3$ are the points at infinity on the three sheets of $\Sigma$), passing through the cut $(a_i^-,a_i^+)$, $(b_j^-,b_j^+)$, and $(c_k^-,c_k^+)$ respectively.

\section{Minimization of $W_{eff}$}

\subsection{Geometric conditions}

We are finally ready to study how the on-shell conditions of the field theory translate  into geometric properties of the Riemann surface $\Sigma$. To do so we need to minimize the effective superpotential \eqref{Weff} as a function of $S_{1,i}$, $S_{2,j}$ and $H_k$. As for the case with a single gauge group and matter in the adjoint and fundamental representations of \cite{CSW3}, it is more convenient to switch to a different set of $2n_1+n_2$ parameters which are directly related to the shape of $\Sigma$. We take as such parameters the coefficients of the polynomials $f_1(z)$, $f_2(z)$ and $g(z)\equiv g_1(z)+g_2(z)$\footnote{We do not consider $g_1(z)$ and $g_2(z)$ separately because, just as for their $\psi^2$ partners, in all physically relevant quantities only their sum appears.} defined in \eqref{polynomials}
\be\label{fg}
\begin{split}
&f_1(z)=\sum_{i=1}^{n_1} f_{1,i}z^{i-1}\\
&f_2(z)=\sum_{j=1}^{n_2} f_{2,j}z^{j-1}\\
&g(z)=\sum_{k=1}^{n_1} g_{k}z^{k-1}
\end{split}
\ee
Equations \eqref{fielddegrees} and \eqref{curve} allow to pass from one set of parameters to the other.
Using \eqref{fielddegrees} we may write in a compact form
\be\label{eqnmoteff}
0=\frac{\partial W_{eff}}{\partial u}=\mathcal{K}\lp\frac{\partial y}{\partial u}~dz\rp
\ee
where $u$ is any of $f_{1,i}$, $f_{2,j}$ or  $g_k$, and $\mathcal{K}$ is an integral operator defined by
\be
\begin{split}
\mathcal{K}&\lp \{N_{1,i}\},\{N_{2,j}\},\{r_k\},\{b_{1,i}\},\{b_{2,j}\},\{h_k\}\rp=\\&\frac{1}{2\pi i}\ln \!\lp\frac{(-\Lambda_0)^{2N_1-N_2}}{\Lambda_1^{2N_1-N_2}}\lp\frac{\Lambda_2}{\Lambda_1} \rp^{\!\!\frac{N_2}{2}}\rp \!\sum_{i=1}^{n_1}\oint_{A_i}-\frac{1}{2\pi i}\ln \!\lp\frac{(-\Lambda_0)^{2N_2-N_1}}{\Lambda_2^{2N_2-N_1}}\lp\frac{\Lambda_1}{\Lambda_2} \rp^{\!\!\frac{N_1}{2}}\rp\!\sum_{j=1}^{n_2}\oint_{B_j}+\\& - \frac{1}{2\pi i} \ln \!\lp\frac{(-\Lambda_0)^{N_1+N_2}}{\Lambda_1^{2N_1-N_2}\Lambda_2^{2N_2-N_1}}\lp\frac{\Lambda_1}{\Lambda_2} \rp^{\!\!\frac{N_1-N_2}{2}}\rp \!  \sum_{k=1}^{n_1}\oint_{C_k}~+\sum_{i=1}^{n_1}N_{1,i}\int_{B_{31}^i} ~+\\&+\sum_{j=1}^{n_2}N_{2,j}\int_{B_{23}^j} ~-\sum_{k=1}^{n_1}r_k\int_{B_{12}^k} ~-\sum_{i=1}^{n_1-1}b_{1,i} \oint_{A_i}~+\sum_{j=1}^{n_2-1}b_{2,j} \oint_{B_j}+~\sum_{k=1}^{n_1}h_k \oint_{C_k}
\end{split}
\ee

In Appendix \ref{holforms1} we show that the one-forms $\frac{\partial y}{\partial u}$ may be written as linear combinations of the $2n_1+n_2-2$ holomorphic one-forms $\omega_{0,k}$, $\omega_{1,k}$ and $\tilde{\omega}_k$ (which make up a basis for the holomorphic one-forms on $\Sigma$), and two meromorphic one-forms with simple poles at points on $\Sigma$ corresponding to $z=\infty$. This allows us to  write $\frac{\partial y}{\partial u}$ as a combination of a more convenient set of holomorphic and meromorphic one-forms.

Let $\xi_i$, $\eta_j$, $\chi_k$ be the basis of holomorphic one-forms on $\Sigma$ defined  as follows
\be\label{performs1}
\begin{split}
&\frac{1}{2\pi i}\oint_{A_l}\xi_i=\delta_{il} \qquad\frac{1}{2\pi i}\oint_{A_{n_1}}\xi_i=-1 \qquad i,l=1,\ldots n_1-1\\
&\frac{1}{2\pi i}\oint_{B_m}\eta_j=\delta_{jm} \qquad\frac{1}{2\pi i}\oint_{B_{n_2}}\eta_j=-1 \qquad j,m=1,\ldots n_2-1\\
&\frac{1}{2\pi i}\oint_{C_p}\chi_k=\delta_{kp} \qquad\frac{1}{2\pi i}\oint_{A_{n_1}}\chi_k=1 \qquad\frac{1}{2\pi i}\oint_{B_{n_2}}\chi_k=-1 \qquad k,p=1,\ldots n_1
\end{split}
\ee
with all other periods vanishing, and let $\tau_{13}$ and $\tau_{23}$ be two meromorphic one-forms with simple poles at infinity, defined by
\begin{gather}
\mathrm{Res} ~\tau_{13}|_{P_1}=-1\qquad \mathrm{Res} ~\tau_{23}|_{P_2}=-1\qquad \mathrm{Res} ~\tau_{13}|_{P_3}=\mathrm{Res} ~\tau_{23}|_{P_3}=1\nonumber\\
\frac{1}{2\pi i}\oint_{A_{n_1}}~\tau_{13}=-1\qquad \frac{1}{2\pi i}\oint_{B_{n_2}}~\tau_{23}=-1 \label{performs2}
\end{gather}
where $P_1$, $P_2$ and $P_3$ are the points corresponding to $z=\infty$ on the three sheets of $\Sigma$ respectively.

We write then
\be
\lp\begin{matrix}\frac{\partial y}{\partial f_{1,i}}\vspace{5pt}\\ \frac{\partial y}{\partial f_{2,j}}\vspace{5pt}\\\frac{\partial y}{\partial g_{k}} \end{matrix}\rp=V \lp\begin{matrix} \xi_i\\\eta_j\\\chi_k\\\tau_{13}\\\tau_{23}\end{matrix}\rp
\ee
where $V$ is an invertible $(2n_1+n_2)\times (2n_1+n_2)$ matrix that will in general depend on $\{f_{1,i}\}$, $\{f_{2,j}\}$ and $\{g_k\}$.

Since $\mathcal{K}$ is an integral operator and $V$ does not depend on either $y$ or $z$, $\mathcal{K}$ and $V$ commute. Thus the equations of motion \eqref{eqnmoteff} become
\be
\mathcal{K}\lp\begin{matrix} \xi_i\\\eta_j\\\chi_k\\\tau_{13}\\\tau_{23}\end{matrix}\rp=0
\ee
which, by using \eqref{performs1} and \eqref{performs2}, read
\be\label{eqnmot}
\begin{split}
&\sum_{i=1}^{n_1}N_{1,i}\int_{B_{31}^i}\xi_l+\sum_{j=1}^{n_2}N_{2,j}\int_{B_{23}^j}\xi_l-\sum_{k=1}^{n_1}r_k\int_{B_{12}^k}\xi_l-2\pi i b_{1,l}=0\\
&\sum_{i=1}^{n_1}N_{1,i}\int_{B_{31}^i}\eta_m+\sum_{j=1}^{n_2}N_{2,j}\int_{B_{23}^j}\eta_m-\sum_{k=1}^{n_1}r_k\int_{B_{12}^k}\eta_m+2\pi i b_{2,m}=0\\
&\sum_{i=1}^{n_1}N_{1,i}\int_{B_{31}^i}\chi_p+\sum_{j=1}^{n_2}N_{2,j}\int_{B_{23}^j}\chi_p-\sum_{k=1}^{n_1}r_k\int_{B_{12}^k}\chi_p+2\pi i h_{p}=0
\end{split}
\ee
\be\nonumber
\begin{split}
&\sum_{i=1}^{n_1}N_{1,i}\int_{B_{31}^i}\!\tau_{13}+\sum_{j=1}^{n_2}N_{2,j}\int_{B_{23}^j}\!\tau_{13}-\sum_{k=1}^{n_1}r_k\int_{B_{12}^k}\!\tau_{13}-\ln \!\lp\frac{(-\Lambda_0)^{2N_1-N_2}}{\Lambda_1^{2N_1-N_2}}\lp\frac{\Lambda_2}{\Lambda_1} \rp^{\!\!\frac{N_2}{2}}\rp =0\\
&\sum_{i=1}^{n_1}N_{1,i}\int_{B_{31}^i}\!\tau_{23}+\sum_{j=1}^{n_2}N_{2,j}\int_{B_{23}^j}\!\tau_{23}-\sum_{k=1}^{n_1}r_k\int_{B_{12}^k}\!\tau_{23}+\ln \!\lp\frac{(-\Lambda_0)^{2N_2-N_1}}{\Lambda_2^{2N_2-N_1}}\lp\frac{\Lambda_1}{\Lambda_2} \rp^{\!\!\frac{N_1}{2}}\rp =0
\end{split}
\ee

\subsection{Compact periods of $\sigma(z)dz$} \label{compact}

We will demonstrate here that the equations of motion \eqref{eqnmot} of the effective superpotential are related to the periods around the cycles $D_i$, $E_j$, $F_k$, and along the non compact paths $B^{n_1}_{31}$ and $B^{n_2}_{12}$ of the meromorphic one-form $\sigma(z)dz$ that is naturally defined on $\Sigma$.

First of all let us express $\sigma(z)dz$ \eqref{sigmaform} in a more convenient way, by making use of the forms \eqref{performs1} and \eqref{performs2}. The boundary conditions \eqref{condsigma} we have imposed on $\sigma(z)dz$ imply that
\be\label{ressigma}
\begin{split}
&\mathrm{Res}~\sigma(z)dz|_{P_1}=-N_1\\
&\mathrm{Res}~\sigma(z)dz|_{P_2}=N_2\\
&\mathrm{Res}~\sigma(z)dz|_{P_3}=N_1-N_2
\end{split}
\ee

Thus from \eqref{condsigma}, \eqref{performs1} and \eqref{performs2} it follows that
\be\label{persigma}
\sigma(z)dz=N_1\tau_{13}-N_2\tau_{23}-\sum_{i=1}^{n_1-1}N_{1,i}\xi_i+\sum_{j=1}^{n_2-1}N_{2,j}\eta_j+\sum_{k=1}^{n_1}r_k\chi_k
\ee

Let us define the $2n_1+n_2-2$ compact cycles (see Figure \ref{figpaths})
\be\label{compactcycles2}
\begin{split}
&D_i\equiv B^i_{31}-B^{n_1}_{31}\qquad i=1\ldots,n_1-1\\
&E_j\equiv B^j_{23}-B^{n_2}_{23}\qquad j=1\ldots,n_2-1\\
&F_k\equiv B^k_{12}+B^{n_2}_{23}+B^{n_1}_{31}\qquad k=1\ldots,n_1
\end{split}
\ee
which have intersection pairings $D_i\cap A_l=\delta_{il}$, $D_i\cap A_{n_1}=-1$, with all other pairings equal to zero, and analogous relations for the pairs $E,B$ and $F,C$.
\FIGURE[!ht]{\epsfig{file=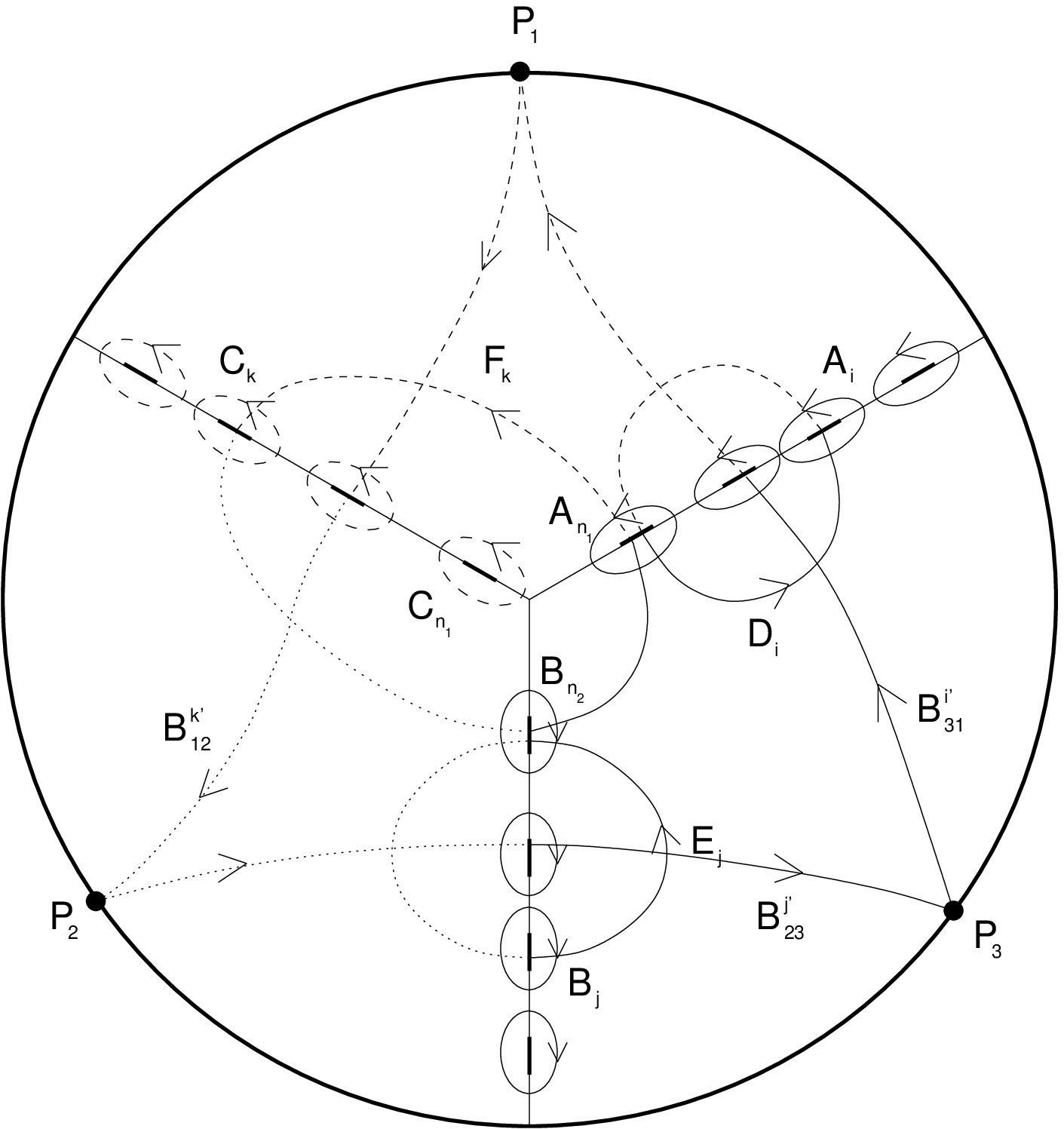,width=0.8\textwidth}\caption{Compact cycles $D_i$, $E_j$ and $F_k$ and non-compact paths connecting points at infinity are represented here. To distinguish between various sheets we have drawn paths on the first sheet with solid lines, while dashed lines represent paths on the second sheet and dotted paths lie on the third sheet.}\label{figpaths}}

The periods of the $1$-form $\sigma(z)dz$ around cycles \eqref{compactcycles2} are easily evaluated via the Riemann bilinear relations of Appendix \ref{AppRiem}. We find
\be\label{sigmacompper}
\begin{split}
&\oint_{D_i} \sigma(z)dz=-\sum_{l=1}^{n_1}N_{1,l}\int_{B_{31}^l} \xi_i -\sum_{j=1}^{n_2}N_{2,j}\int_{B_{23}^j}\xi_i+\sum_{k=1}^{n_1}r_k\int_{B_{12}^k}\xi_i\\
&\oint_{E_j} \sigma(z)dz=\sum_{i=1}^{n_1}N_{1,i}\int_{B_{31}^i} \eta_j +\sum_{l=1}^{n_2}N_{2,l}\int_{B_{23}^l}\eta_j-\sum_{k=1}^{n_1}r_k\int_{B_{12}^k}\eta_j\\
&\oint_{F_k} \sigma(z)dz=-\sum_{i=1}^{n_1}N_{1,i}\int_{B_{31}^i} \chi_k -\sum_{j=1}^{n_2}N_{2,j}\int_{B_{23}^j}\chi_k+\sum_{l=1}^{n_1}r_l\int_{B_{12}^l}\chi_k
\end{split}
\ee

There are still two paths that are interesting to consider because they are non-compact and independent from all the compact cycles we have considered so far. The first of these paths is $B_{31}^{n_1}$ (the integral along which we will write $\int_{P_3}^{P_1}$) which goes from the point $P_3$ at infinity on the third sheet to the corresponding point $P_1$ at infinity on the first sheet, passing through the cut $(a_{n_1}^-,a_{n_1}^+)$. The other one is $B_{23}^{n_2}$ which goes from $P_2$ to $P_3$ passing through $(b_{n_2}^-,b_{n_2}^+)$.

The integrals of $\sigma(z)dz$ along these paths are
\be\label{sigmanoncompper}
\begin{split}
  &\int_{P_3}^{P_1} \sigma(z)dz=\sum_{i=1}^{n_1}N_{1,i}\int_{B_{31}^i}\tau_{13}+\sum_{j=1}^{n_2}N_{2,j}\int_{B_{23}^j}\tau_{13}-\sum_{k=1}^{n_1}r_k\int_{B_{12}^k}\tau_{13}\\
  &\int_{P_2}^{P_3} \sigma(z)dz=-\sum_{i=1}^{n_1}N_{1,i}\int_{B_{31}^i}\tau_{23}-\sum_{j=1}^{n_2}N_{2,j}\int_{B_{23}^j}\tau_{23}+\sum_{k=1}^{n_1}r_k\int_{B_{12}^k}\tau_{23}
\end{split}
\ee

All other non-compact cycles can be written as linear combinations of the $2n_1+n_2$ cycles $\{D_i,E_j,F_k,B_{31}^{n_1},B_{23}^{n_2}\}$ we have considered so far, as can be seen from the relations \eqref{compactcycles2}.

Comparison of \eqref{sigmacompper} and \eqref{sigmanoncompper} with the equations of motion \eqref{eqnmot} for the effective superpotential of the gauge theory, shows that the on-shell condition is translated, in the geometric construction of the theory, into the requirement that the periods of the one-form $\sigma(z) dz$ be integers around the compact cycles $D_i$, $E_j$ and $F_k$. That is
\be\label{geomeqnmot}
\begin{split}
&\frac{1}{2\pi i}\oint_{D_i} \sigma(z)dz=-b_{1,i}\\
&\frac{1}{2\pi i}\oint_{E_j} \sigma(z)dz=-b_{2,j}\\
&\frac{1}{2\pi i}\oint_{F_k} \sigma(z)dz=h_k
\end{split}
\ee

The equations of motion \eqref{eqnmot} also impose conditions on the value of the integrals along the non compact cycles $B_{31}^{n_1}$ and $B_{23}^{n_2}$
\be\label{geomeqnmot2}
\begin{split}
&\int_{P_3}^{P_1} \sigma(z)dz=\ln \!\lp\frac{(-\Lambda_0)^{2N_1-N_2}}{\Lambda_1^{2N_1-N_2}}\lp\frac{\Lambda_2}{\Lambda_1} \rp^{\!\!\frac{N_2}{2}}\rp\\
&\int_{P_2}^{P_3} \sigma(z)dz=\ln \!\lp\frac{(-\Lambda_0)^{2N_2-N_1}}{\Lambda_2^{2N_2-N_1}}\lp\frac{\Lambda_1}{\Lambda_2} \rp^{\!\!\frac{N_1}{2}}\rp
\end{split}
\ee

This result is the generalization to the quiver gauge theory $A_2$ of what has been found in \cite{Ferrari2} and \cite{CSW3} for a gauge theory with a single gauge group and matter in the adjoint, and in the adjoint and fundamental representations respectively.

\subsection{A meromorphic function $\psi(z)$ on $\Sigma$ and the SW curve}

The cycles $A_i$, $B_j$, $C_k$, $D_i$, $E_j$ and $F_k$ with $i=1,\ldots n_1-1$, $j=1,\ldots n_2-1$ and $k=1,\ldots n_1$ form a complete basis for the homology group $H_2(\Sigma)$ of compact cycles on $\Sigma$. With the equations of motion \eqref{geomeqnmot} and the conditions \eqref{condsigma} this is sufficient to ensure that the meromorphic one-form $\sigma(z)dz$ has integer periods along any compact cycle of $\Sigma$. There must exist on $\Sigma$, then, a meromorphic function $\psi(z)$ such that
\be\label{psi}
\sigma(z)dz=d\ln \psi(z)
\ee

From \eqref{ressigma} it follows that $\psi(z)$ must have a pole in $P_1$ of order $N_1$ and zeros in $P_2$ and $P_3$ of order $N_2$ and $N_1-N_2$ respectively\footnote{We have taken $N_1>N_2$ to fix notations, but the results we obtain are easily generalizable also to $N_1\leq N_2$.}.

Since $\psi(z)$ must be defined on $\Sigma$, we may write it as
\be\label{psi2}
\psi(z)=\Pi_1(z)+\chi(z)
\ee
where $\Pi_1(z)$ is a polynomial to be determined, and $\chi(z)$ satisfies
\be\label{chi}
G(\chi,z)=\chi^3+A(z)\chi+B(z)=0
\ee
with $A(z)$ and $B(z)$ polynomials to be determined.

Since $A(z)$ and $B(z)$ are polynomials, they are defined on the complex plane, and won't distinguish between the three sheets of $\Sigma$. In particular each of them will have the same asymptotic behavior at the three points at infinity $P_1$, $P_2$ and $P_3$. Equation \eqref{chi} ensures then that, at infinity, $\chi(z)$ will diverge with the same power law in $P_1$, $P_2$ and $P_3$, but in general with different coefficients. $\Pi_1(z)$ is a polynomial, thus it will also behave in the same way at the three points at infinity. In order to account for the behavior of $\psi(z)$ at infinity, the considerations we just made tell us that both $\Pi_1(z)$ and $\chi(z)$ have degree $N_1$.

Since we want $\chi(z)$ to be defined on $\Sigma$, the surface \eqref{chi} must have the same cut structure as $\Sigma$. The following equation must hold, then
\be\label{samecuts}
D(z)\frac{\partial G}{\partial \chi}(\chi(z),z)=C(z)\frac{\partial F}{\partial y}(y(z),z)
\ee
where $F(y,z)$ is the function that defines $\Sigma$ through \eqref{curve}. $C(z)$ and $D(z)$ are polynomials that account for cuts that have closed up on the two surfaces. In particular the roots of $D(z)=0$ represent cuts of $\Sigma$ that closed up, and $\mathrm{deg}\,D(z)$ equals the sum of the number of vanishing $N_{1,i}$'s, $N_{2,j}$'s and $r_k$'s. 

Equation \eqref{samecuts} has to be interpreted as defined on the complex plane, and not on $\Sigma$. In fact, all cuts appear at the same time, regardless of which sheets they connect. On each sheet, analogous equations can be defined, which account only for the branch points on that sheet. They read
\be\label{samecutsi}
D_i(z)\frac{\partial G}{\partial \chi}(\chi_i(z),z)=C_i(z)\frac{\partial F}{\partial y}(y_i(z),z)
\ee 
where $i=1,2,3$ labels the sheets, and again $C_i(z)$ and $D_i(z)$ count the number of closed cuts on the two surfaces.

Just like $\frac{\partial F}{\partial y}(y_i(z),z)$ (see \eqref{derFinf}), $\frac{\partial G}{\partial \chi}(\chi_i(z),z)$ will in general have different asymptotic behaviors in $P_1$, $P_2$ and $P_3$
\be\label{l}
\frac{\partial G}{\partial \chi}(\chi_i(z),z)=\frac{C_i(z)}{D_i(z)}\frac{\partial F}{\partial y}(y_i(z),z)\sim z^{l_i}
\ee

We assume $l_i< 2N_1$ on all three sheets, which, with $\frac{\partial G}{\partial \chi}(\chi_i(z),z)=3\chi_i^2+A(z)$ and $\mathrm{deg}\,\chi_i=N_1$ for $i=1,2,3$, requires $\mathrm{deg}\,A(z)=2N_1$.

From equation \eqref{samecutsi} it follows that
\be\label{chisqrt}
\chi_i(z)=\sqrt{-\frac{1}{3}A(z)+\frac{C_i(z)}{3D_i(z)}\frac{\partial F}{\partial y}(y_i(z),z)}
\ee

Using equation \eqref{l}, at $P_2$ and $P_3$ equation \eqref{chisqrt} becomes $(j=2,3)$
\be\label{chiP2P3} 
\chi_j(z)\simeq\sqrt{-\frac{1}{3}A(z)}+\frac{1}{2\sqrt{-3A(z)}}\frac{C_j(z)}{D_j(z)}\frac{\partial F}{\partial y}(y_j(z),z)+\ldots=\sqrt{-\frac{1}{3}A(z)}+O\lp z^{l_j-N_1}\rp
\ee

To reproduce the right singular behavior of $\psi(z)$, we must have
\be\label{asympchi}
\begin{split}
&\chi_2(z)=-\Pi_1(z)+O\lp\frac{1}{z^{N_2}}\rp\quad\qquad\quad\;\, \mathrm{in}\;P_2\\
&\chi_3(z)=-\Pi_1(z)+O\lp\frac{1}{z^{N_1-N_2}}\rp\quad\qquad \mathrm{in}\;P_3
\end{split}
\ee
Comparing equations \eqref{asympchi} with \eqref{chiP2P3}, we find
\be\label{sqrtA}
\begin{split}
&\sqrt{-\frac{1}{3}A(z)}=-\Pi_1(z)+O(z^h)\\
&\sqrt{-\frac{1}{3}A(z)}=-\Pi_1(z)+O(z^{\tilde{h}})
\end{split}
\ee
where
\be\label{hhtilde}
h\equiv\max\{l_2-N_1,-N_2\}\qquad  \tilde{h}\equiv\max\{l_3-N_1,-N_1+N_2\}
\ee

Relations \eqref{sqrtA} make sense only if $h=\tilde{h}$. We strengthen our previous assumption on $l_2$ and $l_3$, and require now that $l_2,\,l_3\leq\max\{N_2,N_1-N_2\}$. At the end of the section we will show that, once it is assumed that $l_i< 2N_1$,  this second condition is required by the existence of $\chi(z)$. In order to solve $h=\tilde{h}$ we need to consider three cases: $N_1<2N_2$, $N_1=2N_2$ and $N_1>2N_2$. We find
\be
\begin{split}
N_1<2N_2\quad&\rightarrow\quad\left\{\begin{array}{l}l_2=N_2\\h=\tilde{h}=-N_1+N_2 \end{array}\right. \\
N_1=2N_2\quad&\rightarrow\quad h=\tilde{h}=-N_2\\
N_1>2N_2\quad&\rightarrow\quad\left\{\begin{array}{l}l_3=N_1-N_2\\h=\tilde{h}=-N_2 \end{array}\right.
\end{split}
\ee

Squaring \eqref{sqrtA} we obtain then
\be\label{A}
A(z)=-3\Pi_1^2(z)+\Pi_k(z)
\ee
with $\Pi_k(z)$ a polynomial of degree $k$ given by
\be\label{k}
k=\left\{\begin{array}{lll}N_2 &\quad & N_1<2N_2\\N_2=N_1-N_2 & & N_1=2N_2\\N_1-N_2 & & N_1>2N_2\end{array}\right.
\ee

Equation \eqref{chi} can now be written
\be\label{chi2}
\chi^3+(-3\Pi_1^2(z)+\Pi_k(z))\chi+B(z)=0
\ee
where the polynomial $B(z)$ is still to be determined. To this purpose we consider this last relation at points $P_2$ and $P_3$. Substituting the first of \eqref{asympchi} into \eqref{chi2} we find
\be\label{BP2}
B(z)=-2\Pi_1^3(z)+\Pi_1(z)\Pi_k(z)+O\lp z^{N_1-2N_2}\rp+O\lp z^{k-N_2}\rp
\ee
while substituting the second of \eqref{asympchi} into \eqref{chi2} we have
\be\label{BP3}
B(z)=-2\Pi_1^3(z)+\Pi_1(z)\Pi_k(z)+O\lp z^{-N_1+2N_2}\rp+O\lp z^{k-N_1+N_2}\rp
\ee

We now have to distinguish among the three cases in \eqref{k}. If $N_1<2N_2$ then $k=N_2$ and equations \eqref{BP2} and \eqref{BP3} may be rewritten as
\be\label{Bmin}
\begin{split}
&B(z)=-2\Pi_1^3(z)+\Pi_1(z)\Pi_k(z)+O\lp 1 \rp\\
&B(z)=-2\Pi_1^3(z)+\Pi_1(z)\Pi_k(z)+O\lp z^{2N_2-N_1}\rp
\end{split}
\ee
where we have neglected all negative-power contributions since $B(z)$ is a polynomial.

The two conditions \eqref{Bmin} must be satisfied simultaneously since they represent the expression of $B(z)$ at two points which are distinct on the three-fold surface \eqref{chi}, but are coincident on the $z$-plane. Being $B(z)$ a polynomial it is defined on the complex plane and does not distinguish between different sheets. Thus
\be\label{Bfix}
B(z)=-2\Pi_1^3(z)+\Pi_1(z)\Pi_k(z)+\gamma
\ee
where $\gamma$ is a non-vanishing constant and all higher contributions in the second of \eqref{Bmin} must cancel, fixing the asymptotic behavior of $\chi(z)$ while it approaches $P_3$.

A similar analysis holds also for the remaining two cases in \eqref{k}, giving the same result as in \eqref{Bfix}, with the vanishing condition of all higher order in the first of \eqref{Bmin} (in the case $N_1>2N_2$) fixing the behavior of $\chi(z)$ near $P_2$.

Finally $\chi(z)$ is defined by
\be\label{chifix}
\chi^3+\lp-3\Pi_1^2(z)+\Pi_k(z)\rp\chi-2\Pi_1^3(z)+\Pi_1(z)\Pi_k(z)+\gamma=0
\ee

As a check we show that $\psi(z)$ as defined in \eqref{psi2} has no zeros at finite $z$. Let $\bar{z}$ be a point in the complex plane such that
\be
\psi(\bar{z})=0
\ee
which, from \eqref{psi2}, requires $\chi(\bar{z})=-\Pi(\bar{z})$. Substituting
this into \eqref{chifix}, all terms cancel but $\gamma$, giving the condition $\gamma=0$ for the existence of a finite zero of $\psi(z)$, which is clearly never satisfied.

This check also justifies the assumption we made after \eqref{hhtilde} on the asymptotic behavior of $\frac{C_j(z)}{D_j(z)}\frac{\partial F}{\partial y}$. Were our assumption on $l_2$, $l_3$ in \eqref{l} not satisfied, we would have $h=\tilde{h}=l_2-N_1=l_3-N_1$ and equation \eqref{Bfix} would have to be modified by substituting the non-vanishing constant $\gamma$ with a non-vanishing polynomial of degree $l_2-\max\{N_2,N_1-N_2\}$, but this would allow for the existence of finite zeros of $\psi(z)$.

At this point we only need to determine $\Pi_1(z)$, $\Pi_k(z)$ and $\gamma$. This is achieved by solving equation \eqref{psi} with respect to $\Pi_1(z)$, $\Pi_k(z)$ and  $\gamma$. The left-hand side of \eqref{psi} is given by \eqref{sigmaform}, where $l_1(z)$, $l_2(z)$ and $r(z)$ are fixed by imposing \eqref{condsigma}.

From \eqref{samecutsi} we write
\be\label{factor}
\begin{split}
&3\chi^2(y_i(z),z)-3\Pi_1^2(z)+\Pi_k(z)=H(y_i(z),z)C_i(z)\\
&3y_i^2(z)-p_2(z)=H(y_i(z),z)D_i(z)
\end{split}
\ee

Once equation \eqref{psi} is solved, we use the first equation in \eqref{factor} to determine $H(y_i(z),z)$ on each sheet, plug it into the second one, and use the three equations (one for each sheet of $\Sigma$) we obtain to determine the unknown parameters $f_1(z)$, $f_2(z)$ and $g(z)$ of the curve $\Sigma$.

The surface we obtain is the on-shell algebraic curve associated with the $\mathcal{N}=1$ $A_2$ quiver gauge theory with superpotential $W_{tree}$ \eqref{Wtree}.

A last point deserves some comments. What does the algebraic curve \eqref{chifix} represent? To answer this question let us use \eqref{psi2} and \eqref{chifix} to determine the surface described by $\psi(z)$. We find
\be\label{SW}
\psi^3-3\Pi_1(z)\psi^2+\Pi_k(z)\psi+\gamma=0
\ee
which has exactly the same form as the Seiberg-Witten curve of the $\mathcal{N}=2$ $A_2$ quiver gauge theory \cite{{Klemm:1996bj},{SW}}.

The relationship between the curve in \eqref{SW} and the SW curve is even deeper. As in \cite{Cach}, let us take the superpotentials $W_1(z)$ and $W_2(z)$ to be of degree $N_1+1$ and $N_2+1$ respectively. In general they will have $N_1$ and $N_2$ distinct critical points. Let us take $N_1<2N_2$ so that both gauge groups are asymptotically free, and consider the Coulomb branch of the theory. We place one eigenvalue of $\Phi_1$ and $\Phi_2$ in each critical point of $W_1(z)$ and $W_2(z)$ respectively. We have then $N_{1,i}=1$ and $N_{2,j}=1$ for all $i$'s and $j$'s, and $r_k=0$ for all $k$'s. All Higgs cuts are actually closed and $\mathrm{deg}\,D(z)$ in \eqref{samecuts} is equal to $N_1$. We now use \eqref{samecuts} to show that for this choice of vacuum the polynomial $C(z)$ must be a constant. The functions $\frac{\partial F}{\partial y}$ and $\frac{\partial G}{\partial \chi}$ read
\be
\begin{split}
&\lp\frac{\partial F}{\partial y}(y(z),z)\rp^2 \sim {W_1^\prime}^2(z){W_2^\prime}^2(z)\lp W_1^\prime(z)+W_2^\prime(z)\rp^2+\ldots\\
&\lp\frac{\partial G}{\partial \chi}(\chi(z),z)\rp^2\sim \Pi_1^2(z)\Pi_k^2(z)+\ldots
\end{split}
\ee
where dots represent lower degree polynomials. Matching the degrees of the polynomials on the left-hand side and right-hand side of \eqref{samecuts}, requires that 
\be\label{degC}
\mathrm{deg}\,C(z)=0
\ee

Equation \eqref{factor} tells us that, in this case, the curve $\psi(z)$ is everywhere regular, and is isomorphic to the surface $H(y,z)$.

With the choice of superpotentials we made, the genus of $\Sigma$ is $2N_1+N_2-2$. From \eqref{samecuts} and the considerations we just made, the genus of the curve \eqref{SW} must be $N_1+N_2-2$, which is what is expected for the SW curve of a $U(N_1)\times U(N_2)$ theory.

Since $z$ has dimension of a mass, being it a coordinate on the moduli space of the theory, dimensional analysis requires that
\be
\begin{split}
&\Pi_k(z)=\Lambda^{2N_1-N_2}\tilde{\Pi}_k(z)\\
&\gamma=\tilde{\Lambda}^{3N_1}
\end{split}\ee
where $\Lambda$ and $\tilde{\Lambda}$ are  dimension-one combinations of the energy scales $\Lambda_1$ and $\Lambda_2$ of the theory, and $\tilde{\Pi}_k(z)$ is a polynomial of degree $N_2$ with a dimensionless coefficient for the highest degree term.

Now, let us consider the classical limit of \eqref{samecuts}. Taking $\Lambda_1$ and $\Lambda_2$ to zero, we find
\be\label{samecutsSW}
{W_1^\prime}(z){W_2^\prime}(z)\lp W_1^\prime(z)+W_2^\prime(z)\rp=9\alpha D(z)\Pi_1(z) \tilde{\Pi}_k(z)
\ee
where the constant $C$ was used to cancel the factor $\Lambda^{2N_1-N_2}$ on the right-hand side up to a dimensionless multiplicative constant $\alpha$. 

Since we left all Higgs cuts empty, the factor $\lp W_1^\prime(z)+W_2^\prime(z)\rp$ will be proportional to $D(z)$, and the most natural interpretation of \eqref{samecutsSW} is
\be\label{proportional}
\begin{split}
&W_1^\prime(z)\sim \Pi_1(z)\\
&W_2^\prime(z)\sim \tilde{\Pi}_k(z)
\end{split}
\ee

The critical points of $W_1(z)$ and $W_2(z)$ single out a point in the $(N_1+N_2)$-dimensional classical moduli space of the $\mathcal{N}=2$ theory. Equation \eqref{proportional} tells us that classically this is the same point as that singled out by the roots of $\Pi_1(z)$ and $\tilde{\Pi}_k(z)$. Again this is what we expect from the SW curve. We argue, then, that \eqref{SW} is the SW curve of the $\mathcal{N}=2$ theory.

The determination of $\Pi_1(z)$, $\Pi_k(z)$ and $\gamma$ through \eqref{psi}, would allow also to describe of the quantum deformed moduli space of the theory.

What happens if we change the superpotentials? The construction of $\psi(z)$ shows that the form of the cubic equation it satisfies, depends only on the asymptotic behavior we require from it, and not on the particular form of the superpotentials. What the superpotentials do is, actually, selecting a point on the moduli space. When we choose superpotentials of degree $N_1+1$ and $N_2+1$, they will single out a generic point in the moduli space, where  the SW curve is, in general, regular. This agrees with the fact that, for such superpotentials, the degree of $C(z)$ in \eqref{factor} is zero (see \eqref{degC}), and the curve $\psi(z)$ is regular.
On the contrary, when we choose superpotentials of degrees smaller than $N_1+1$ and $N_2+1$, they  will single out a point where some monopoles are massless in the $\mathcal{N}=2$ theory \cite{{Cach}, {Ferrari2}, {CSW3}, {CSW2}} and the SW curve will be singular. But again this agrees with equation \eqref{factor}. Since we stay in the Coulomb branch the degree of $D(z)$ is still $N_1$, $\frac{\partial G}{\partial \chi}$ does not depend on the superpotentials and does not change, and the left-hand side of \eqref{samecutsi} is unchanged. But  $\frac{\partial F}{\partial y}$ is of a smaller degree than it was in the previous case, and for \eqref{samecutsi} to hold we must have $\mathrm{deg}\,C(z)>0$. Equation \eqref{factor} tells us that $\psi(z)$ is now singular.

The SW curve for the special case $N_1=N_2$ and its derivation via matrix model techniques, have also been studied in \cite{naculich2} using a different approach.

\section{Conclusions}

In this paper we have studied the $\mathcal{N}=1$ $A_2$ quiver gauge theory using matrix models and a generalized Konishi anomaly. We used the loop equations for the resolvents $R_1(z)$ and $R_2(z)$ of the gauge theory to associate a non-hyperhelliptic curve $\Sigma$ to the theory \cite{quiver}
\be
y^3- p_2(z)y - p_3(z)=0
\ee
and obtained the relations satisfied by $w_a^\alpha(z)$ and $T_a(z)$ defined in \eqref{operators}, which with the resolvents determine the low-energy dynamics of the theory.

The request that the vacuum be supersymmetric requires $w_a^\alpha(z)$ to vanish identically, and the form of the solution to the simplified equations for $T_a(z)$ made it natural to define on $\Sigma$ a one-form $\sigma(z)dz$. The periods of this form around the $A$-cycles of $\Sigma$ (which in the paper we labeled $A_i$, $B_j$ and $C_k$) are fixed by the choice of the classical vacuum (e.g. by the gauge group breaking pattern \eqref{breakc} or \eqref{breakh}).

The geometric description of the Coulomb and Higgs phases of this theory is very similar to that of the $U(N)$ theory with matter in the adjoint representation only \cite{{Ferrari2}, {CSW2}}. In this last case  the branch points of the elliptic surface associated with the theory coincide with the critical points of the superpotential, which are the classically allowed eigenvalues of the adjoint superfield. When such eigenvalues are placed at one of these branch points a cut opens up, corresponding to a hole in the surface. In the $A_2$ theory the Coulomb and Higgs phases are distinguished by the constraints that must be satisfied by the eigenvalues of the superfields $\Phi_1$ and $\Phi_2$ (and by the expectation value of the bifundamental hypermultiplets), but again these correspond to the conditions fulfilled by the branch points of $\Sigma$. In our case also, when eigenvalues are placed at one of these (double) branch points, a hole in $\Sigma$ opens up connecting two of the three sheets. Thus we argue that different phases with the same low-energy $U(1)^k$ gauge group are continuously connected as in \cite{{Ferrari2}, {CSW2}}, while jumps in the number of $U(1)$ factors are only possible through a singular phase transition (that in the $\mathcal{N}=2$ theory corresponds to a monopole becoming massless).

The surface $\Sigma$, as dealt with so far, is actually the curve associated with the off-shell theory. We have demonstrated that the minimization of the effective superpotential can be translated into a geometric language: the one-form $\sigma(z)dz$ must have integer periods also around the $B$-cycles of $\Sigma$ (which we labeled $D_i$, $E_j$ and $F_k$). The geometric translation of the on-shell conditions was first introduced in \cite{{Ferrari2}, {CSW3}} for theories with a single $U(N)$ gauge group.

The meromorphic one-form $\sigma(z)dz$ has integer periods around all compact cycles of the on-shell non-hyperelliptic curve $\Sigma$. This ensures that there exists on $\Sigma$ a meromorphic function whose logarithm  $\sigma(z)dz$ is the differential. We have found this function and showed that it determines a surface
\be
\psi^3-3\Pi_1(z)\psi^2+\Lambda^{2N_1-k}\tilde{\Pi}_k(z)\psi+\tilde{\Lambda}^{3N_1}=0
\ee
which we have argued to be the Seiberg-Witten curve for the $\mathcal{N}=2$ theory.

\acknowledgments
We would like to thank Alberto Zaffaroni for useful discussions and suggestions. This work was partially supported by INFN and MURST under contract 2001-025492, and by the European Commission TMR program HPRN-CT-2000-00131, in association to the University of Padova.

\appendix
\section{Holomorphic $1$-forms on $\Sigma$}\label{holforms}

The genus $g$ of the Riemann surface $\Sigma$ \eqref{curve} can be evaluated through the formula
\be
2g-2=n(2g^\prime-2)+\sum_{branch~ points}(\nu_P-1)
\ee
where $g^\prime$ is the genus of the surface of which $\Sigma$ is the $n$-fold cover (in our case the covered surface is the complex plane, thus $g^\prime=0$) and $\nu_P$ is the number of sheets that meet at the branch point $P$. We obtain then that the genus of $\Sigma$ is $2n_1+n_2-2$ and that on $\Sigma$ there exist $2n_1+n_2-2$ holomorphic one-forms. Let us consider the one-forms
\be\label{forms}
\omega_R(y,z)=\frac{R(y,z)}{\frac{\partial F}{\partial y}(y,z)}~dz=\frac{a(z)y^2+b(z)y+c(z)}{3y^2-p_2(z)}~dz
\ee
where $a(z)$, $b(z)$ and $c(z)$ are polynomials. We only consider up to quadratic terms in $y$ because \eqref{curve} relates any higher degree term to up to quadratic terms. We look for  holomorphic forms among \eqref{forms}. Because of the way they have been written, it is clear that all of \eqref{forms} are holomorphic in a neighborhood of any point of $\Sigma$ except for the points at infinity and the branch points, which we will now consider.

We have to distinguish between the points at infinity. We have that, because of \eqref{curve}, $y\sim z^{n_1}$ for $P\rightarrow P_1,P_2,P_3$, but the behavior of $3y^2-p_2(z)$ differs according to which point at infinity we are approaching. When considering the surface at infinity, that is far away from the branch points and the cuts, it actually makes no difference whether we are considering the exact quantum-deformed curve or its classical limit, the asymptotic behavior is unchanged. Thus near the points at infinity we have \eqref{roots}
\be\label{sheetinf}
\begin{split}
&y_1(z)\simeq -\frac{2}{3}W_1^\prime(z)-\frac{1}{3}W_2^\prime(z)\\
&y_2(z)\simeq \frac{1}{3}W_1^\prime(z)+\frac{2}{3}W_2^\prime(z)\\
&y_3(z)\simeq \frac{1}{3}W_1^\prime(z)-\frac{1}{3}W_2^\prime(z)
\end{split}
\ee
while, being a regular function, $p_2(z)$ always goes like
\be
3 p_2(z)\simeq {W_1^\prime}(z)^2+W_1^\prime(z) W_2^\prime (z)+{W_2^\prime}(z)^2
\ee
Then we have
\be\label{derFinf}
\begin{split}
&3y(z)^2-p_2(z)\sim z^{2n_1}\qquad\quad (y,z)\rightarrow P_1\\
&3y(z)^2-p_2(z)\sim z^{n_1+n_2}\qquad (y,z)\rightarrow P_2\\
&3y(z)^2-p_2(z)\sim z^{n_1+n_2}\qquad (y,z)\rightarrow P_3
\end{split}
\ee

Let
\be\label{omega2}
\omega_{2,k}=\frac{z^k y^2}{3y^2-p_2(z)}~dz
\ee
A good chart at infinity is $z=\frac{1}{u}$ and $y=\frac{1}{u^{n_1}}$, with $u\rightarrow 0$. We have
\be\begin{split}
&\omega_{2,k}\sim \frac{u^{-k-2n_1}}{u^-2n_1}\frac{du}{u^2}=u^{-k-2}du \qquad\qquad\quad\,\, (y,z)\rightarrow P_1\\
&\omega_{2,k}\sim \frac{u^{-k-2n_1}}{u^-n_1-n_2}\frac{du}{u^2}=u^{n_2-n_1-2-k}du \qquad (y,z)\rightarrow P_2,P_3\\
\end{split}\ee
which make $\omega_{2,k}$ holomorphic at infinity only for $k\leq -2$. Forms with negative $k$ are not holomorphic in $z=0$, then there are no holomorphic $\omega_{2,k}$ on the whole surface $\Sigma$.

We consider next
\be\label{omega1}
\omega_{1,k}=\frac{z^k y}{3y^2-p_2(z)}~dz
\ee
which at infinity behaves as
\be\label{inf1}\begin{split}
&\omega_{1,k}\sim \frac{u^{-k-n_1}}{u^-2n_1}\frac{du}{u^2}=u^{n_1-2-k}du \qquad\quad\,\,\,(y,z)\rightarrow P_1\\
&\omega_{1,k}\sim \frac{u^{-k-n_1}}{u^-n_1-n_2}\frac{du}{u^2}=u^{n_2-k-2}du \qquad (y,z)\rightarrow P_2,P_3\\
\end{split}
\ee
thus $\omega_{1,k}$ is holomorphic at infinity for $k\leq n_2-2$.

We also  have to check the behavior of  $\omega_{1,k}$ around the branch points. It is easy to show that all of these forms are holomorphic around such points. At the branch points $z$ is not a good coordinate to describe the surface, but $y$ is a good choice. Because of \eqref{implicit}
\be
\begin{split}
&z-z_{br.pt.}\sim (y-y_{br.pt.})^2\\
&3y(z)-p_2(z)\sim y-y_{br.pt.}
\end{split}\ee
for $(y,z)\rightarrow$ branch point, and $\omega_{1,k}$ is holomorphic around any branch point for $k\geq 0$.

Thus  $\omega_{1,k}$ is holomorphic on $\Sigma$ for $0\leq k \leq n_2-2$.

Eventually we consider the one-forms
\be\label{omega0}
\omega_{0,k}=\frac{z^k }{3y^2-p_2(z)}~dz
\ee

The same argument that showed that the forms $\omega_{1,k}$ are holomorphic around any branch point of $\Sigma$, ensures that also all $\omega_{0,k}$ with $k\geq 0$ are holomorphic in a neighborhood of each branch point.

At infinity we have
\be\label{inf2}\begin{split}
&\omega_{0,k}\sim \frac{u^{-k}}{u^-2n_1}\frac{du}{u^2}=u^{2n_1-2-k}du \qquad\quad\quad\,\,\,\,(y,z)\rightarrow P_1\\
&\omega_{0,k}\sim \frac{u^{-k}}{u^-n_1-n_2}\frac{du}{u^2}=u^{n_1+n_2-k-2}du \qquad (y,z)\rightarrow P_2,P_3\\
\end{split}
\ee
thus $\omega_{0,k}$ is holomorphic on $\Sigma$ for $0\leq k \leq n_1+n_2-2$.

We have found $n_1+2n_2-2$ holomorphic forms on $\Sigma$, we are still missing $n_1-n_2$. By analyzing \eqref{inf1} and \eqref{inf2}, we see that there are forms which are holomorphic around $P_1$ but not around $P_2$ and $P_3$, and that there are exactly $n_1-n_2$ pairs of  such forms among the $\omega_{1,k}$'s and $\omega_{0,k}$'s. This is  suggesting that for every $n_2-2\leq k \leq n_1-2$  there might be a combination of $\omega_{1,k}$ and $\omega_{0,k}$ which is holomorphic also around $P_2$ and $P_3$. We found this is indeed the case.

The combination we should take can be guessed by looking at \eqref{sheetinf}. Since the singularities were on the second and third sheets we take
\be\label{omegatilde}
\tilde{\omega}_k=\frac{z^{n_2-2+k} \lp y(z)-\frac{1}{3}\sum_{p=n_1+2-k}^{n_1+1}g_{1,p}z^{p-1}\rp }{\frac{\partial F}{\partial y}(y,z)}~dz \qquad\quad 1\leq k \leq n_1-n_2
\ee
where $g_{1,p}$ are the coefficients of $W_1(\Phi_1)$ \eqref{chpotential}.

Obviously such forms are holomorphic around the  branch points of $\Sigma$ and $P_1$ because the single terms are. We need to check their behavior only around $P_2$ and $P_3$
\be
\tilde{\omega}_k\sim \frac{u^{-n_2+2-k}~u^{-n_1+k}}{u^{-n_1-n_2}}\frac{du}{u^2}
\ee

The forms $\tilde{\omega}_k$ are holomorphic on $\Sigma$ for $1\leq k\leq n_1-n_2$. This completes the list of holomorphic one-forms that can be built on $\Sigma$.

\subsection{The one-forms $\frac{\partial y}{\partial u}$} \label{holforms1}

We show in this subsection how the forms $\frac{\partial y}{\partial f_{1,i}}dz$, $\frac{\partial y}{\partial f_{2,j}}dz$ and $\frac{\partial y}{\partial g_{k}}dz$ may be written as linear combinations of the holomorphic one-forms found in the first part of this appendix and two meromorphic forms with simple poles at infinity.

We derive equation \eqref{curve} with respect to $u$, where again $u$ stands for any of $f_{1,i}$, $f_{2,j}$ or $g_{k}$. We find
\be
\frac{\partial F(y,z)}{\partial y}\frac{\partial y}{\partial u}-y\frac{\partial p_2(z)}{\partial u}-\frac{\partial p_3(z)}{\partial u}=0
\ee
from which
\be
\frac{\partial y}{\partial u}=\frac{\partial p_2(z)}{\partial u}\frac{y}{\frac{\partial F(y,z)}{\partial y}}+\frac{\partial p_3(z)}{\partial u}\frac{1}{\frac{\partial F(y,z)}{\partial y}}
\ee

These resemble very closely the holomorphic one-forms we built in \eqref{omega1}, \eqref{omega0} and \eqref{omegatilde}. In fact from \eqref{chpotential}, \eqref{p2p3} and \eqref{fg}, we may write
\be
\begin{split}
\frac{\partial y}{\partial f_{1,i}}(z)dz=&\frac{1}{4}\omega_{1,i-1}-\frac{1}{12}\sum_{p=1}^{n_1+1}g_{1,p}\omega_{0,i+p-2}-\frac{1}{6}\sum_{p=1}^{n_2+1}g_{2,p}\omega_{0,i+p-2}\qquad 1\leq i\leq n_2-1\\
\frac{\partial y}{\partial f_{1,i}}(z)dz=&\frac{1}{4}\tilde{\omega}_{i-n_2+1}-\frac{1}{12}\sum_{p=1}^{n_1+n_2-i}\!\!\!\!g_{1,p}\omega_{0,i+p-2}-\frac{1}{6}\sum_{p=1}^{n_2+1}g_{2,p}\omega_{0,i+p-2}\quad n_2\leq i\leq n_1-1\\
\frac{\partial y}{\partial f_{1,n_1}}(z)dz=&\frac{1}{4}\frac{z^{n_1-1}\lp y-\frac{1}{3}\sum_{p=n_2+1}^{n_1+1}g_{1,p}z^{p-1}\rp}{\frac{\partial F(y,z)}{\partial y}}dz-\frac{1}{6}g_{2,n+1}\frac{z^{n_1+n_2-1}}{\frac{\partial F(y,z)}{\partial y}}dz+\\&-\frac{1}{12}\sum_{p=1}^{n_2}(g_{1,p}+2g_{2,p})\omega_{0,n_1+p-2}\\
\frac{\partial y}{\partial f_{2,j}}(z)dz=&\frac{1}{4}\omega_{1,j-1}+\frac{1}{6}\sum_{p=1}^{n_1+1}g_{1,p}\omega_{0,j+p-2}+\frac{1}{12}\sum_{p=1}^{n_2+1}g_{2,p}\omega_{0,j+p-2}\qquad 1\leq j\leq n_2-1\\
\frac{\partial y}{\partial f_{2,n_2}}(z)dz=&\frac{1}{4}\tilde{\omega}_1+\frac{1}{12}\sum_{p=1}^{n_2+1}g_{2,p}\omega_{0,p+n_2-2}+\frac{1}{6}\sum_{p=1}^{n_1}g_{1,p}\omega_{0,p+n_2-2}+\frac{g_{1,n_1+1}z^{n_1+n_2-1}}{4~\frac{\partial F(y,z)}{\partial y}}dz\\
\frac{\partial y}{\partial g_k}dz=&\frac{1}{4}\omega_{0,k-1}
\end{split}
\ee

We only need to demonstrate that the two meromorphic one-forms that appeared in the previous equations have simple poles at infinity and are locally holomorphic everywhere else. Because of the considerations of the previous part of this appendix, this last point needs no more discussion, thus we are left with analyzing the behavior of the two one-forms at the three points at infinity.

We have
\be
\begin{split}
&\frac{z^{n_1-1}\lp y-\frac{1}{3}\sum_{p=n_2+1}^{n_1+1}g_{1,p}z^{p-1}\rp}{\frac{\partial F(y,z)}{\partial y}}dz\sim \frac{u^{-n_1+1}u^{-n_1}u^{-2}}{u^{-2n_1}}~du=\frac{1}{u}~du\qquad(y,z)\rightarrow P_1\\
&\frac{z^{n_1-1}\lp y-\frac{1}{3}\sum_{p=n_2+1}^{n_1+1}g_{1,p}z^{p-1}\rp}{\frac{\partial F(y,z)}{\partial y}}dz\sim \frac{u^{-n_1+1}u^{-n_2}u^{-2}}{u^{-n_1-n_2}}~du=\frac{1}{u}~du\qquad (y,z)\rightarrow P_2,P_3\\
\end{split}
\ee
and
\be
\begin{split}
&\frac{z^{n_1+n_2-1}}{\frac{\partial F}{\partial y}(y,z)}dz\sim \frac{u^{-n_1-n_2+1}u^{-2}}{u^{-2n_1}}~du=u^{n_1-n_2-1}~du\qquad(y,z)\rightarrow P_1\\
&\frac{z^{n_1+n_2-1}}{\frac{\partial F}{\partial y}(y,z)}dz\sim \frac{u^{-n_1-n_2+1}u^{-2}}{u^{-n_1-n_2}}~du=\frac{1}{u}~du\qquad\quad\qquad(y,z)\rightarrow P_2,P_3
\end{split}
\ee

\section{Computations in the matrix model}\label{appmatcomp}

In this appendix we evaluate the derivative of the free energy of the matrix model with respect to the low energy degrees of freedom $S_{1,i}$ and $S_{2,j}$\footnote{This computation is similar to the one performed in \cite{CSW3}. It presents, though, some more complications which make reporting it worthwhile.}. Since equation \eqref{freeen} is valid for any $A_n$ quiver theory, we will keep this generality as far as possible, and specialize to the $A_2$ case only when it will be necessary.

First of all we have to write $\mathcal{F}_0$ as a function of the $S_{a,i}$'s. To do so, we introduce a set of chemical potentials. Eq. \eqref{freeen} becomes
\be
\begin{split}
\mathcal{F}_0=& \sum_{k=1}^{n} \lp \int \mathrm{d}x \rho_k(x) W_k(x) -\int \mathrm{d}x ~\mathrm{d}x^\prime \rho_k(x)\rho_k(x^\prime) \ln \frac{\arrowvert x-x^\prime \arrowvert}{\Lambda_k}  \rp +\\&+\sum_{k=1}^{n-1}   \int \mathrm{d}x ~\mathrm{d}x^\prime \rho_k(x)\rho_{k+1}(x^\prime) \ln \frac{\arrowvert x-x^\prime \arrowvert}{\sqrt{\Lambda_k \Lambda_{k+1}}}-\sum_{a=1}^n\sum_{i=1}^{n_a}\mu_{a,i}\hat{S}_{a,i}
\end{split}
\ee
where
\be
\hat{S}_{a,i}=\int_{a_{a,i}^-}^{a_{a,i}^+}d\lambda ~\rho_a(\lambda)=\frac{1}{2\pi i}\oint_{A_{a,i}}dz~\omega_{a}(z)
\ee

Let $\lambda$ be any eigenvalue of $\hat{\Phi}_a$ belonging to the $i^{th}$ cut $(a_{a,i}^-,a_{a,i}^+)$, then the equation of motion for the eigenvalue density $\rho_a(\lambda)$ is
\be\label{freneqmot}
\begin{split}
0=\frac{\delta\mathcal{F}_0}{\delta\rho_a(\lambda)}=&W_a(\lambda)-2\int dx ~\rho_a(x)\ln\frac{|\lambda-x|}{\Lambda_a}+\int dx~\rho_{a+1}(x)\ln\frac{|\lambda-x|}{\sqrt{\Lambda_a\Lambda_{a+1}}}+\\&+\int dx~\rho_{a-1}(x)\ln\frac{|\lambda-x|}{\sqrt{\Lambda_{a-1}\Lambda_a}}-\mu_{a,i}
\end{split}\ee
which is also valid for $a=1$ and $a=n$ as long as we fix $\rho_{-1}(\lambda)=\rho_{n+1}(\lambda)=0$. This equation allows us to evaluate the eigenvalue density $<\rho_a(\lambda)>$. Once we have fixed $\rho_a(\lambda)$, we can write the free energy as a function of the chemical potentials, from which we derive the free energy as a function of the $S_{a,i}$ via a Legendre transform
\be
\begin{split}
&\mathcal{F}_0(\{\mu_{a,i}\})=\mathcal{F}_0(\{\mu_{a,i}\};<\rho_a(\lambda)>)\\
&\hat{S}_{a,i}=-\frac{\partial\mathcal{F}_0(\{\mu_{b,j}\})}{\partial\mu_{a,i}}\\
&\mathcal{F}_0(\{\hat{S}_{a,i}\})=\mathcal{F}_0(\{\mu_{a,i}\})+\sum_{a=1}^n\sum_{i=1}^{n_a}\mu_{a,i}\hat{S}_{a,i}
\end{split}
\ee

We can now evaluate the quantity we are interested in
\be\label{derF0}
\begin{split}
\frac{\partial\mathcal{F}_0(\{\hat{S}_{b,j}\})}{\partial \hat{S}_{a,i}}=&\mu_{a,i}=~W_a(\lambda)+\frac{1}{2\pi i}\sum_{j=1}^{n_{a-1}}\oint_{A_{a-1,j}}\!\!\!\!\!\!\!\!\!\!dz~\omega_{a-1}(z)\ln\frac{z-\lambda}{\sqrt{\Lambda_{a-1}\Lambda_a}}+\\&-\frac{2}{2\pi i}\sum_{j=1}^{n_{a}}\oint_{A_{a,j}}\!\!\!\!\!\!dz~\omega_{a}(z)\ln\frac{z-\lambda}{\Lambda_a}+\frac{1}{2\pi i}\sum_{j=1}^{n_{a+1}}\oint_{A_{a+1,j}}\!\!\!\!\!\!\!\!\!\!dz~\omega_{a+1}(z)\ln\frac{z-\lambda}{\sqrt{\Lambda_a\Lambda_{a+1}}}
\end{split}
\ee
where we have used \eqref{freneqmot} and \eqref{dens-res}.

Since $\mu_{a,i}$ is $\lambda$-independent, the whole right-hand side of \eqref{derF0} does not depend on $\lambda$. We can thus choose $\lambda$ to be any eigenvalue on the $(a,i)^{th}$ cut. As in \cite{CSW3} we choose for convenience $\lambda=a_{a,i}^+$.

We now specialize to the case of an $A_2$ quiver. For both $a=1,2$ in \eqref{derF0} there are two kinds of cycles $A_{a,l}$, those associated with Coulomb branch points (where $W_a^\prime(z)=0$) and those associated with the Higgs branch points (where $ W_1^\prime(z)+W_2^\prime(z)=0$). In the main text we called $A_i$ and $B_j$ the Coulomb cycles for $a=1$ and $a=2$ respectively, and $C_k$ the Higgs cycles.

We have then
\be
\begin{split}
&\int dx ~\rho_1(x)=\frac{1}{2\pi i}\sum_{i=1}^{n_1}\oint_{\tilde{A}_i}dz~\omega_1(z)+\frac{1}{2\pi i}\sum_{k=1}^{n_1}\oint _{C_k} dz ~\omega_1(z)\\
&\int dx ~\rho_2(x)=\frac{1}{2\pi i}\sum_{j=1}^{n_2}\oint_{\tilde{B}_j}dz~\omega_2(z)+\frac{1}{2\pi i}\sum_{k=1}^{n_1}\oint _{\tilde{C}_k} dz ~\omega_2(z)
\end{split}
\ee

We first consider $\lambda$ on the cut $(a_i^-,a_i^+)$. We may write \eqref{derF0} as
\be\label{interm1}
\begin{split}
\frac{\partial \mathcal{F}_0}{\partial \hat{S}_{1,i}}=&W_1(a_i^+)+\frac{1}{2\pi i}\sum_{j=1}^{n_2}\oint_{\tilde{B}_j} dz~\omega_2(z) \ln \frac{z-a_i^+}{\sqrt{\Lambda_1\Lambda_2}}+\frac{1}{2\pi i}\sum_{k=1}^{n_1}\oint_{\tilde{C}_k} dz~\omega_2(z) \ln \frac{z-a_i^+}{\sqrt{\Lambda_1\Lambda_2}}+\\&-\frac{2}{2\pi i}\sum_{l=1}^{n_1}\oint_{\tilde{A}_l} dz~\omega_1(z) \ln \frac{z-a_i^+}{\Lambda_1}-\frac{2}{2\pi i}\sum_{k=1}^{n_1}\oint_{C_k} dz~\omega_1(z) \ln \frac{z-a_i^+}{\Lambda_1}
\end{split}
\ee

We take the cut of the logarithm to run from $a_i^+$ to the cut-off $\Lambda_0$. We now deform the contours in \eqref{interm1}. In doing so we have to be careful and perform the deformation on the right sheet. The resolvent  $\omega_1(z)$ is defined on the first sheet of $\Sigma$, and we find
\be\label{deformA}
\sum_{i=1}^{n_1}\oint_{\tilde{A}_i}=\int_{a_i^+}^{\Lambda_0}+\int_{\Lambda_0}^{e^{2\pi i} \Lambda_0}+\int_{\Lambda_0}^{a_i^+}-\sum_{k=1}^{n_1}\oint_{C_k}
\ee
while $\omega_2(z)$ is defined on the second sheet of $\Sigma$, and we have
\be\label{deformB}
\sum_{j=1}^{n_2}\oint_{\tilde{B}_j}=\int_{a_i^+}^{\Lambda_0}+\int_{\Lambda_0}^{e^{2\pi i} \Lambda_0}+\int_{\Lambda_0}^{a_i^+}-\sum_{k=1}^{n_1}\oint_{\tilde{C}_k}
\ee

After a little complex analysis, and using eqns. \eqref{degrees} and \eqref{H1=H2}, \eqref{interm1} may be written as (at first order in $1/\Lambda_0$)
\be\label{interm2}
\begin{split}
\frac{\partial \mathcal{F}_0}{\partial \hat{S}_{1,i}}=~&W_1(a_i^+)+\hat{S}_2\ln\lp-\frac{\Lambda_0}{\sqrt{\Lambda_1\Lambda_2}}\rp-2\hat{S}_1\ln\lp-\frac{\Lambda_0}{\Lambda_1}\rp-\hat{H}\ln\lp-\frac{\Lambda_0}{\Lambda_1}\sqrt{\frac{\Lambda_2}{\Lambda_1}}\rp+\\&+\int_{a_i^+}^{\Lambda_0}\!\!\!\!dz\lp\omega_1(z)-\omega_2(z)\rp+\int_{a_i^+}^{\Lambda_0}\!\!\!\!dz~\omega_1(z)
\end{split}
\ee

A similar analysis leads to
\be\label{interm3}
\begin{split}
\frac{\partial \mathcal{F}_0}{\partial \hat{S}_{2,j}}=~&W_2(b_j^+)\!+\hat{S}_1\ln\lp-\frac{\Lambda_0}{\sqrt{\Lambda_1\Lambda_2}}\rp-2\hat{S}_2\ln\lp-\frac{\Lambda_0}{\Lambda_2}\rp-\hat{H}\ln\lp-\frac{\Lambda_0}{\Lambda_2}\sqrt{\frac{\Lambda_1}{\Lambda_2}}\rp+\\&+\int_{b_j^+}^{\Lambda_0}\!\!\!\!dz\lp\omega_2(z)-\omega_1(z)\rp+\int_{b_j^+}^{\Lambda_0}\!\!\!\!dz~\omega_2(z)
\end{split}
\ee

The computation of the derivative with respect to $\hat{H}_k$ requires a little more care because of \eqref{H1=H2}. We first treat $\hat{H}_{1,k}$ and $\hat{H}_{2,k}$ as independent variables, and use \eqref{derF0} to evaluate the derivative of the free energy with respect to them
\be
\begin{split}
\frac{\partial \mathcal{F}_0}{\partial\hat{ H}_{1,k}}=~&W_1(c_k^+)-2\hat{S}_1\ln\lp-\frac{\Lambda_0}{\Lambda_1}\rp+\hat{S}_2\ln\lp-\frac{\Lambda_0}{\sqrt{\Lambda_1\Lambda_2}}\rp-\hat{H}\ln\lp-\frac{\Lambda_0}{\Lambda_1}\sqrt{\frac{\Lambda_2}{\Lambda_1}}\rp+\\&-\int_{c_k^+}^{\Lambda_0}\!\!\!\!dz~\omega_2(z)+2\int_{c_k^+}^{\Lambda_0}\!\!\!\!dz~\omega_1(z)\\
\frac{\partial \mathcal{F}_0}{\partial\hat{ H}_{2,k}}=~&W_2(c_k^+)-2\hat{S}_2\ln\lp-\frac{\Lambda_0}{\Lambda_2}\rp+\hat{S}_1\ln\lp-\frac{\Lambda_0}{\sqrt{\Lambda_1\Lambda_2}}\rp-\hat{H}\ln\lp-\frac{\Lambda_0}{\Lambda_2}\sqrt{\frac{\Lambda_1}{\Lambda_2}}\rp+\\&-\int_{c_k^+}^{\Lambda_0}\!\!\!\!dz~\omega_1(z)+2\int_{c_k^+}^{\Lambda_0}\!\!\!\!dz~\omega_2(z)
\end{split}
\ee
Then equation \eqref{H1=H2} tells us that $\hat{H}_{1,k}$ and $\hat{H}_{2,k}$ are actually the same degree of freedom, and the meaningful quantity to consider is the derivative of the free energy with respect to $\hat{H}_k\equiv \hat{H}_{1,k}=\hat{H}_{2,k}$. At this point, this is easily computed by taking the sum of the derivatives we have just evaluated. We obtain
\be\label{interm4}
\begin{split}
\frac{\partial \mathcal{F}_0}{\partial\hat{ H}_{k}}=\frac{\partial \mathcal{F}_0}{\partial\hat{ H}_{1,k}}+\frac{\partial \mathcal{F}_0}{\partial\hat{ H}_{2,k}}=~W_1(c_k^+)+W_2(c_k^+)+\int_{c_k^+}^{\Lambda_0}\!\!\!\!dz~\omega_1(z)+\int_{c_k^+}^{\Lambda_0}\!\!\!\!dz~\omega_2(z)+&\\-\hat{S}_1\ln\lp-\frac{\Lambda_0}{\Lambda_1}\sqrt{\frac{\Lambda_2}{\Lambda_1}}\rp-\hat{S}_2\ln\lp -\frac{\Lambda_0}{\Lambda_2}\sqrt{\frac{\Lambda_1}{\Lambda_2}}\rp-2\hat{H}\ln\lp-\frac{\Lambda_0}{\sqrt{\Lambda_1\Lambda_2}}\rp&
\end{split}\ee

We are not yet satisfied with the form of eqs. \eqref{interm2}, \eqref{interm3}, \eqref{interm4}: we want to express them in terms of geometrical data, instead of referring to the matrix model resolvents. To do so, we need to use relations  expressing  matrix model resolvents in terms of coordinates on the algebraic surface $\Sigma$ and derivatives of the superpotentials. In \cite{quiver} we found such relations for the gauge theory, which we reported in \eqref{roots}. Because of the dictionary between matrix models and gauge theories, they are valid also for the matrix model resolvents
\be\label{res/sheets}
\begin{split}
\omega_1(z)&=y_1(z)+\frac{2}{3}W_1^\prime(z)+\frac{1}{3}W_2^\prime(z)\\
\omega_2(z)&=-y_2(z)+\frac{1}{3}W_1^\prime(z)+\frac{2}{3}W_2^\prime(z)\\
\omega_1(z)-\omega_2(z)&=-y_3(z)+\frac{1}{3}W_1^\prime(z)-\frac{1}{3}W_2^\prime(z)\\
\end{split}
\ee

Making use of \eqref{res/sheets} we rewrite \eqref{interm2}, \eqref{interm3} and \eqref{interm4} as (up to terms of order $1/\Lambda_0$ or higher)
\be\label{derF0fin}
\begin{split}
\frac{\partial \mathcal{F}_0}{\partial \hat{S}_{1,i}}=&W_1(\Lambda_0)+\hat{S}_2\ln\!\!\lp\frac{-\Lambda_0}{\sqrt{\Lambda_1\Lambda_2}}\rp-2\hat{S}_1\ln\!\!\lp-\frac{\Lambda_0}{\Lambda_1}\rp-\hat{H}\ln\!\!\lp-\frac{\Lambda_0}{\Lambda_1}\sqrt{\frac{\Lambda_2}{\Lambda_1}}\rp+\int_{B^i_{31}}\!\!\!\!dz~ y(z)\\
\frac{\partial \mathcal{F}_0}{\partial \hat{S}_{2,j}}=&W_2(\Lambda_0)+\hat{S}_1\ln\!\!\lp\frac{-\Lambda_0}{\sqrt{\Lambda_1\Lambda_2}}\rp-2\hat{S}_2\ln\!\!\lp-\frac{\Lambda_0}{\Lambda_2}\rp-\hat{H}\ln\!\!\lp-\frac{\Lambda_0}{\Lambda_2}\sqrt{\frac{\Lambda_1}{\Lambda_2}}\rp+\!\int_{B^j_{23}}\!\!\!\!dz~ y(z)\\
\frac{\partial \mathcal{F}_0}{\partial \hat{H}_{k}}=&W_1(\Lambda_0)+W_2(\Lambda_0)-\hat{S}_1\ln\lp-\frac{\Lambda_0}{\Lambda_1}\sqrt{\frac{\Lambda_2}{\Lambda_1}}\rp-\hat{S}_2\ln\lp -\frac{\Lambda_0}{\Lambda_2}\sqrt{\frac{\Lambda_1}{\Lambda_2}}\rp+\\&-2\hat{H}\ln\lp-\frac{\Lambda_0}{\sqrt{\Lambda_1\Lambda_2}}\rp-\int_{B^k_{12}}\!\!\!\!dz~ y(z)
\end{split}
\ee
where the paths $B_{31}^{i}$, $B_{23}^{j}$ and $B_{12}^k$ were defined in section \ref{matrixpot} and figure \ref{figpaths}.


\section{Riemann bilinear relations on $\Sigma$}\label{AppRiem}


We report here a list of Riemann bilinear relations which we used in section \ref{compact}. They can all be obtained from \eqref{performs1}, \eqref{performs2} and
\be
2\pi i\lp\mathrm{Res} \,\mathcal{G}\rp|_R\int_S^R\mathcal{H}-2\pi i\lp\mathrm{Res} \,\mathcal{H}\rp|_T\int_Q^T\mathcal{G}=\pm\sum_{l=1}^g\lp \oint_{\hat{A}_l}\mathcal{H}\oint_{\hat{B}_l}\mathcal{G}- \oint_{\hat{A}_l}\mathcal{G}\oint_{\hat{B}_l}\mathcal{H}\rp
\ee
where $\mathcal{G}$ and $\mathcal{H}$ are two one-forms with simple poles in $R,S$ and $Q,T$ respectively, and $g$ is the genus of the surface. The sign of the right-hand side of the equation depends on the orientation of the cycles. In our case $g=2n_1+n_2-2$ and the cycles $\hat{A}_l$ and $\hat{B}_l$ are given by
\be
\begin{split}
\hat{A}_i\rightarrow A_i\qquad \hat{B}_i\rightarrow D_i\qquad i=1,\ldots n_1-1\\
\hat{A}_{j+n_1-1}\rightarrow B_j\qquad \hat{B}_{j+n_1-1}\rightarrow E_j\qquad j=1,\ldots n_2-1\\
\hat{A}_{k+n_1+n_2-2}\rightarrow C_k\qquad \hat{B}_{k+n_1+n_2-2}\rightarrow F_k\qquad k=1,\ldots n_1
\end{split}
\ee

We find
\be
\begin{split}
&\int_{P_3}^{P_1}\xi_i=-\oint_{D_i}\tau_{13} \qquad\int_{P_3}^{P_1}\eta_j=\oint_{E_j}\tau_{13} \qquad\int_{P_3}^{P_1}\chi_k=-\oint_{F_k}\tau_{13}\\
&\int_{P_2}^{P_3}\xi_i=\oint_{D_i}\tau_{23} \qquad\int_{P_2}^{P_3}\eta_j=-\oint_{E_j}\tau_{23} \qquad\int_{P_2}^{P_3}\chi_k=\oint_{F_k}\tau_{23}\\
&\int_{P_1}^{P_2}\xi_i=\oint_{D_i}\tau_{12} \qquad\int_{P_1}^{P_2}\eta_j=-\oint_{E_j}\tau_{12} \qquad\int_{P_1}^{P_2}\chi_k=-\oint_{F_k}\tau_{12}
\end{split}
\ee
\be\nonumber
\begin{split}
&\oint_{D_i}\xi_l=\oint_{D_l}\xi_i\qquad \oint_{D_i}\eta_j=-\oint_{E_j}\xi_i\qquad \oint_{D_i}\chi_k=\oint_{F_k}\xi_i\\
&\oint_{E_j}\eta_l=\oint_{E_l}\eta_j\qquad\oint_{E_j}\chi_k=-\oint_{F_k}\eta_j\qquad \oint_{F_k}\chi_l=\oint_{F_l}\chi_k\\
&\int_{P_2}^{P_3}\tau_{13}=-\int_{P_3}^{P_1}\tau_{23}\qquad\int_{P_3}^{P_1}\tau_{12}=-\int_{P_1}^{P_2}\tau_{13}\qquad\int_{P_2}^{P_3}\tau_{12}=\int_{P_1}^{P_2}\tau_{23}
\end{split}
\ee

\end{document}